\documentclass[fleqn,usenatbib]{mnras}
\usepackage{amssymb}

\usepackage[toc,page]{appendix}
\usepackage{amssymb}

\usepackage[T1]{fontenc}
\usepackage{ae,aecompl}
\usepackage{graphicx, subfigure}

\usepackage{graphicx}	
\usepackage{amsmath}	
\usepackage{amssymb}	

\usepackage{commath}

\newcommand{\comment}[1]{}

\newcommand{\sims}{{\textsc{Teahupoo} }}
\newcommand{\simsborder}{{\textsc{Teahupoo}}}

\title[Flux estimation with deep learning]{The PAU survey: Estimating galaxy photometry with deep learning}

\author[Cabayol et al.]{L.~Cabayol$^{1}$\thanks{E-mail:lcabayol@ifae.es},
M.~Eriksen$^{1}$\thanks{E-mail: eriksen@pic.es} \thanks{Also at Port d'Informaci\'{o} Cient\'{i}fica (PIC), Campus UAB, C. Albareda s/n, 08193 Bellaterra (Cerdanyola del Vall\`{e}s), Spain},
A.~Amara$^{4}$,
J.~Carretero$^{1}$\footnotemark[3],
R.~Casas$^{2,3}$, 
\newauthor
F.~J.~Castander$^{2,3}$,
J.~De.~Vicente$^{8}$,
E.~Fern\'andez$^{1}$, 
J.~Garc\'ia-Bellido$^{5}$,
\newauthor
E.~Gaztanaga$^{2,3}$,
H.~Hildebrandt$^{6}$,
R.~Miquel$^{1,7}$, 
C.~Padilla$^{1}$,
\newauthor
E.~S\'anchez$^{8}$,
S.~Serrano$^{2}$,
I.~Sevilla-Noarbe$^{8}$, 
P.~Tallada-Cresp\'i$^{8}$\footnotemark[3] 
\\ \\
$^{1}$Institut de F\'{\i}sica d'Altes Energies (IFAE), The Barcelona Institute of Science and Technology, 08193 Bellaterra (Barcelona), Spain \\
$^{2}$Institute of Space Sciences (ICE, CSIC),  Campus UAB, Carrer de Can Magrans, s/n,  08193 Barcelona, Spain\\
$^{3}$Institut d'Estudis Espacials de Catalunya (IEEC), 08193 Barcelona, Spain\\
$^{4}$Institute of Cosmology \& Gravitation, University of Portsmouth, Dennis Sciama Building, Burnaby Road, Portsmouth PO1 3FX, UK\\
$^{5}$Instituto de Fisica Teorica UAM/CSIC, Universidad Autonoma de Madrid, 28049 Madrid, Spain\\
$^{6}$Ruhr University Bochum, Faculty of Physics and Astronomy, Astronomical Institute (AIRUB), \\
German Centre for Cosmological Lensing, 44780 Bochum, Germany\\
$^{7}$Instituci\'o Catalana de Recerca i Estudis Avan\c{c}ats, E-08010 Barcelona, Spain\\
$^{8}$Centro de Investigaciones Energ\'eticas, Medioambientales y Tecnol\'ogicas (CIEMAT), Madrid, Spain\\
}

\date{Accepted XXX. Received YYY; in original form ZZZ}

\pubyear{}

\begin{document}
\label{firstpage}
\pagerange{\pageref{firstpage}--\pageref{lastpage}}
\maketitle
\begin{abstract}
With the dramatic rise in high-quality galaxy data expected from {\it Euclid} and Vera C. Rubin Observatory, there will be increasing demand for fast high-precision methods for measuring galaxy fluxes. These will be essential for inferring the redshifts of the galaxies. In this paper, we introduce \textsc{Lumos}, a deep learning method to measure photometry from galaxy images. \textsc{Lumos} builds on \textsc{BKGnet}, an algorithm to predict the background and its associated error, and predicts the background-subtracted flux probability density function. We have developed \textsc{Lumos} for
data from the Physics of the Accelerating Universe Survey (PAUS), an imaging survey using a 40 narrow-band filter camera (PAUCam). PAUCam images are affected by scattered light, displaying a background noise pattern that can be predicted and corrected for. On average, \textsc{Lumos} increases the SNR of the observations by a factor of 2 compared to an aperture photometry algorithm. It also incorporates other advantages like robustness towards distorting artefacts, e.g. cosmic rays or scattered light, the ability of deblending and less sensitivity to uncertainties in the galaxy profile parameters used to infer the photometry. Indeed, the number of flagged photometry outlier observations is reduced from 10\% to 2\%, comparing to aperture photometry. Furthermore, with \textsc{Lumos} photometry, the photo-z scatter is reduced by $\approx$10\% with the Deepz machine learning photo-z code and the photo-z outlier rate by 20\%. The photo-z improvement is lower than expected from the SNR increment, however currently the photometric calibration and outliers in the photometry seem to be its limiting factor.
\end{abstract}



\section{Introduction}
Wide field galaxy surveys are a powerful tool for cosmology. Galaxy redshifts are the most fundamental property of
any cosmological or galaxy evolution study. Spectroscopic
surveys, e.g. the Sloan Digital Sky Survey \citep[SDSS,][]{SDSS}, measure very high precision redshifts, however
these are only possible for on the order of a million objects \citep[e.g. BOSS,][]{boss}. In contrast, imaging surveys are $\approx$ 2 orders of magnitude ahead in terms of
number of objects but these are observed with a lower
spectral resolution, which makes the redshift measurements
less precise. Current and past imaging surveys, e.g. The
Kilo-Degree Survey \citep[KiDS,][]{Kids}, The Hyper
Supreme-Cam Subaru \citep[HSC,][]{hsc} or The
Dark Energy Survey \citep[DES,][]{DES} have
detected hundreds of millions of galaxies and oncoming
surveys like Euclid \citep[][]{Euclid} or Vera C. Rubin
Observatory \citep[LSST,][]{LSST} will increase this
number to billions. Consequently, fast and precise
methods to analyse and extract galaxy properties (e.g. 
flux, size, shape) are needed.\\

There are many different algorithms to estimate galaxy photometry. One  very widely used example is \textsc{SExtractor} \citep{SExtractor}, which applies aperture photometry \citep{forced_photometry} inspired by the \cite{Kron} first moments algorithm. This technique measures the flux of the targeted galaxies by placing an aperture around the source and measuring the light captured inside such aperture. Another technique is model fitting \citep{modellfitting}, which consists of fitting the galaxy image to a theoretical model and extract its photometry. This includes the \textsc{GaaP} \citep{GaaP} algorithm, which estimates the total flux by fitting the  pixelated galaxy images to polar shapelets \citep{shapelets,shapelets2}, separating the galaxy image into components with explicit rotational symmetries.\\

There are many other examples as e.g. \textsc{ProFound} \citep{profound}, \textsc{T-PHOT} \citep{tphot} and \textsc{Tractor} \citep{tracotr}, and each of them is adjusted to outperform the others on a particular data set. For instance, a photometry algorithm can be optimised to work very well on images with many blended galaxies \citep[e.g.][]{deblending1} while another can be intended to improve the photometry of very noisy galaxies. Therefore, depending on the type of data and the science goals, different methodologies are applied to improve the photometry estimation. \\

Although all these algorithms have proven to work well, they also have their shortcomings. Aperture photometry works very well on clean images, but it is not robust towards distorting effects such as e.g. blended galaxies, variable background light and cosmic rays. On the other hand, model fitting is sensitive to model parametrisation.  In contrast, this is not the case for machine learning techniques, which learn a model adapted to the data. Furthermore, deep learning has proven to be very powerful on image recognition and computer vision \citep[e.g.][]{dl_objdec,dl_objdet2}, which makes it a robust tool to work on images with artifacts and variant effects. Also, the evaluation of a trained machine learning algorithm is very fast, which is relevant when dealing with very large amounts of data. For instance, \cite{tools_comp} compares several source-extraction codes and conclude that currently there is no tool  sufficiently fast and accurate to be well suited to large-scale automated segmentation. \\

Deep learning has already been applied to different steps of astronomical imaging photometry, e.g. photometry of blended galaxies \citep{deblending1}, PSF simulation \citep{PSFsims}, cosmic ray rejection \citep{deepCR} or source detection  \citep[][]{galaxy_detection_dl}. The power of deep learning techniques on object detection or image recognition tasks makes these steps of the data reduction, among others, very suitable candidates to apply machine learning. While addressing them with classical methods can be difficult and computationally expensive, deep learning is an effective tool to tackle the problem. \\

In this paper, we present \textsc{Lumos}\footnote{The code is available at https://github.com/PAU-survey/lumos under a GPL-3 license.}, a deep learning based algorithm to extract the photometry from astronomical images. It consists of a Convolutional Neural Network that works on input galaxy images and a Mixture density network that outputs the probability distribution of galaxy fluxes. \textsc{Lumos} builds on \textsc{BKGnet} \citep{bkgnet}, a deep learning algorithm to predict and correct strongly varying astronomical backgrounds. \textsc{Lumos} estimates the probability distribution of the background subtracted galaxy flux, which requires the implicit estimation and subtraction of the background noise.\\

We have developed and tested \textsc{Lumos} using  images from the Physics of the Accelerating Universe  Survey (PAUS). PAUS is an imaging survey that measures  high precision photo-z to faint magnitudes $i_{\rm AB}<22.5$,  while covering a large area of sky \citep{photoz-Marti}. This is possible thanks to the PAUCam instrument \citep{PAUCAM_Francisco, PAUCam_Padilla_2016,PAUCam_Padilla_2019},  a camera equipped with 40 narrow band filters covering the optical spectrum \citep{Casas2016}.  With \textsc{BCNz2} \citep{photoz-Martin}, a template fitting algorithm, PAUS reaches a photo-z precision $\sigma(z)/(1+z)\approx 0.0035$ for the best 50\% of the sample, compared to typical precision of 0.05 for broad band measurements. Similar results are obtained with \textsc{Delight}, a hybrid template-machine-learning photometric redshift algorithm that uses Gaussian processes \citep{Delight}. Both methods are further improved with \textsc{Deepz} \citep{Deepz}, a deep learning algorithm to measure photo-zs that reduces the $\sigma_{\rm 68}$ scatter by 50\% compared with the template fitting method, having the largest improvement at fainter galaxies. Furthermore, \citealt{alex_photozs}  presents a high precision photo-z catalogue in the COSMOS field  computed using a combination of PAUS NB and 26  broad, intermediate, and narrow bands covering the UV, visible and near infrared spectrum. Although \textsc{Lumos} has been  developed for PAUS, it can be adapted to any imaging survey, like e.g. {\it Euclid}  or Rubin.\\

The structure of this paper is as follows. In $\S$\ref{sec:data}, we  present the PAUS data used in  this paper and the simulations we use for training. Section \ref{sec:methods}  presents different flux estimation alternatives that we have used to compare to \textsc{Lumos} performance. In \S\ref{sec:lumos_general}, we introduce \textsc{Lumos}, its architecture and the training procedure. Section \ref{sec:resultssims} presents \textsc{Lumos} results on simulations, including validation of the flux probability distributions, a comparison with alternative flux estimation methods and deblending tests. Finally, $\S$\ref{sec:pausresults} shows \textsc{Lumos} results on the PAUS data, including single exposure photometry, co-added fluxes and photometric redshifts obtained with \textsc{Lumos} photometry. Conclusions and discussion can be found in \S\ref{sec:conclusions}.
 
\section{Data}
\label{sec:data}
In this section, we present PAUS data ($\S$\ref{sec:pausdata}) and \sims simulations ($\S$\ref{sec:sims_astropy}), the simulated galaxy images used throughout the paper.

\subsection{PAUS data}
\label{sec:pausdata}
PAUS data are taken with the William Herschel Telescope (WHT), at the Observatorio del Roque de los Muchachos in La Palma at the Canary Islands (Spain). Images are obtained  with the PAUCam instrument \citep{PAUCAM_Francisco,PAUCam_Padilla_2019}, an optical camera equipped with 40 narrow band filters (NB), covering a wavelength range from 4500 to 8500\AA\ \citep{Casas2016}. The NB filters have 130\AA\ full width half maximum (FWHM) and a separation between consecutive bands of 100\AA. The camera has 18 red-sensitive fully depleted Hamamatsu CCD detectors \citep{Casas2012}, although only the 8 central CCDs are currently used for NB imaging. Each CCD has 4096x2048 pixels with a pixel scale of 0.263 arcsec/pix. The NB filters are mounted in five trays with 8 filters per tray that can be exchanged and placed in front of the central CCDs. The NB filter set effectively measures a low resolution spectrum ($R\approx 50$). \\

PAUS has been observing since the 2015B semester and as of 2021A, PAUS has taken data for 160 nights. It partially covers the  CFHTLS fields\footnote{http://www.cfht.hawaii.edu/Science/CFHTLS\_Y\_WIRCam\\/cfhtlsdeepwidefields.html} W1, W2 and W3 and the COSMOS field\footnote{http://cosmos.astro.caltech.edu/}. Currently, PAUS data has a 40 narrow band coverage of  10 deg$^2$ in W1 and W2,  20 deg$^2$ in W3 and the 2 deg$^2$ of the COSMOS field. The PAUS data are stored at the Port d'Informaci\'o Cient\'ifica (PIC), where the data are processed and distributed \citep{Tonello}.\\
 
\subsubsection{COSMOS sample}
This paper focuses on data from the COSMOS field \citep[][]{cosmos_field}, which were taken in the semesters 2015B, 2016A, 2016B and 2017B. These are the data used in the \textsc{BKGnet} paper and also in all the photo-z studies published so far. The COSMOS field observations comprise a total of 9749 images, 243 images in each NB. While observing COSMOS, the PAU camera was modified to mitigate the effect of scattered light by introducing baffles on the edges of the NB filters of each filter tray. This camera intervention changed the noise patterns of PAUS images. Half of the images in COSMOS were taken before the camera modification and the other half, after. The baseline exposure times in the COSMOS field are 70, 80, 90, 110 and 130 seconds from the bluest to the reddest filter trays. The complete photometry catalogue comprises 64,476 galaxies to $i_{\rm AB}$ < 23 in 40 NB filters. This corresponds to  $\approx$ 12,5 million galaxy observations ($\approx$ 5 observations per galaxy and NB filter). \\
 
\subsubsection{Standard data reduction pipeline}
 PAUS data reduction process consists of two pipelines: the \textsc{Nightly} and the \textsc{MEMBA} pipelines. The \textsc{Nightly} pipeline (Serrano et al. in prep.) is the first step and performs an instrumental de-trending processing, e.g. electronic or illumination biases. Electronic biases are corrected with an overscan subtraction, an observation with the shutter closed and zero exposure time. Vignetting of the telescope corrector, pixel-to-pixel variations or dead pixels are detected and corrected with dome flats, 10 second exposures of a uniformly and homogeneously  illuminated screen.
 The Point Spread Function (PSF) is modeled with \textsc{PSFex} \citep{psfex} using star observations from the COSMOS Advanced Camera for Surveys \citep[ACS,][]{acs}. Cosmic rays are detected and masked from the image using a Laplacian edge detector \citep{CR}. The  astrometry of the narrow band images is calibrated using \textsc{SCAMP} \citep{scamp} with the  \textit{Gaia} DR2 stars \citep{GaiaDR2} as a reference catalogue. \\
 
The \textsc{MEMBA} pipeline (Serrano et al. in prep.) is the second step in the data reduction. It applies forced aperture photometry to targets selected from an external detection catalogue (see $\S$\ref{sec:methods:aperture_phot} for  more detailes). In the COSMOS field, the detection parent catalogue is provided by \cite{Ilbert2008cat} and the photometry calibration is relative to SDSS stars (Castander et al. in prep.). A brief description of the photometry calibration can be found in \cite{photoz-Martin}.\\

The machine learning algorithm developed in this paper presents an alternative to \textsc{MEMBA}. It aims to be more robust in the presence of distorting effects as blending or scattered light. It also aims to be provide better photometry measurements of galaxies with errors in the parent catalogue galaxy profile parameters.

\subsection{\sims simulations }
\label{sec:sims_astropy}

\begin{figure}
\includegraphics[width= 0.48\textwidth]{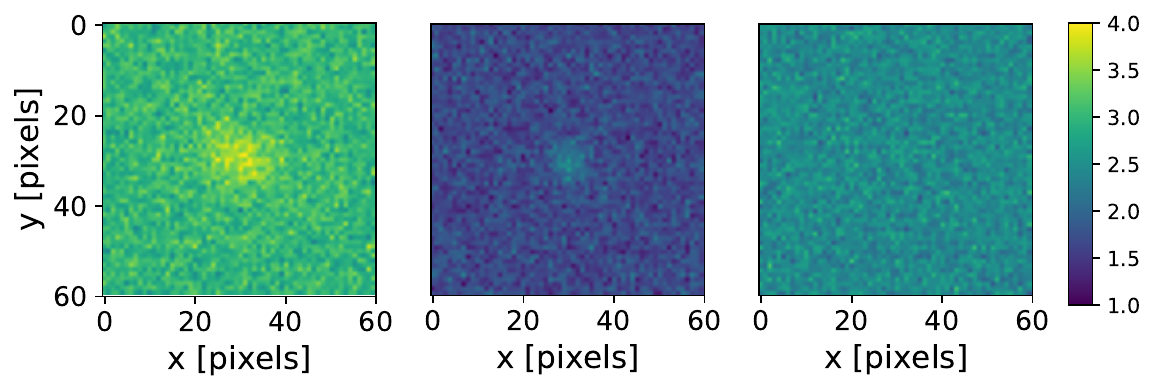}
\centering
\caption{From left to right, \sims galaxy images with $i$-band magnitudes  $18.5, 20.3$ and $22.3$, in PAUS "NB685". These are simulated with a exposure time of 90 seconds,  the baseline PAUS exposure time in the "NB685" filter.}
\label{fig:sims_astropy}
\end{figure}

For this work, we have constructed the  \simsborder\footnote{Named after the favourite sandwich of the first author.} simulations, a set of PAUS-like galaxy image simulations. Three examples of \sims  galaxies with $i$-band magnitudes $i_{\rm AB} = 18.5, 20.3$ and 22.3 are shown in Figure \ref{fig:sims_astropy}.  Note that already at $i_{\rm AB} \approx 20$ it is hard to distinguish the galaxy from the background noise and  with $i_{\rm AB} > 22$, the galaxy signal is visually masked by background fluctuations. \\

\begin{figure}
\includegraphics[width=0.48\textwidth]{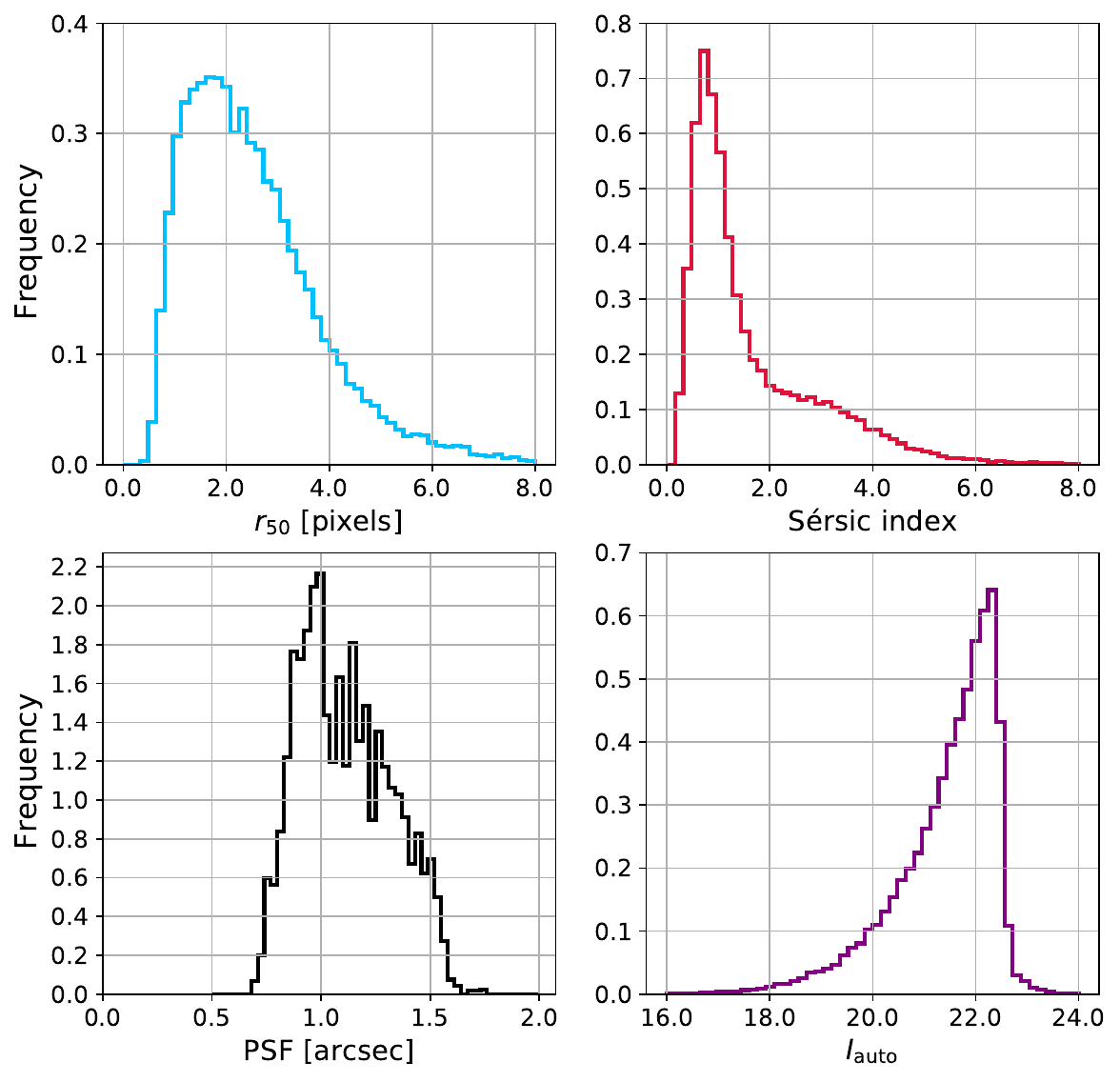}
\centering
\caption{ Distributions of the half light radius ($r_{\rm 50}$) (top left), S\'{e}rsic index ($n_{\rm s}$) (top right), the PSF FWHM (bottom left)  and the $I_{\rm auto}$ magnitude of PAUS galaxies in the COSMOS field (bottom right).}
\label{fig:PAUSsims}
\end{figure}

\sims light profiles are modelled with a single S\'{e}rsic profile. Each galaxy is generated using
\begin{equation}
    I(R) = I_{\rm e}\exp{\left\{ -(2n_{\rm s} -1/3) \left[ \left( \frac{R}{R_{\rm 50}}\right)^{\rm 1/n_{\rm s}}-1\right]\right \}},
    \label{eq:sersic_sims}
\end{equation} where $I_{\rm e}$, $r_{\rm 50}$ and $n_{\rm s}$ are the the surface brightness, the half light radius and the S\'{e}rsic index, respectively.  As the shape and the size of the galaxy are correlated, to ensure galaxies are realistic, we jointly sample the half light radius, the S\'{e}rsic index and the ellipticity from their distributions in the COSMOS field. Elliptical galaxies are simulated by elongating the half light radius according to the $b/a$ distribution in the COSMOS field. Figure \ref{fig:PAUSsims} shows the distributions of $r_{\rm 50}$ (top left) and the S\'{e}rsic index (top right) of PAUS galaxies in the COSMOS field, which are provided by \cite{Ilbert2008}.\\

\sims image simulations are 60x60 pixels image generated with \textsc{Astropy} \citep{astropy:2013,astropy:2018}.
\textsc{Astropy} methods evaluate the galaxy profiles (Eq. \ref{eq:sersic_sims}) at the center of each pixel instead of integrating along the pixel. This is problematic for small and steep galaxy profiles (high $n_{\rm s}$), where the flux changes significantly along the pixel. To correct for this effect, galaxies are initially drawn in a 600x600 grid with a later size reduction. Furthermore, drawing on a larger grid allows shifting the galaxy at a sub-pixel level from the center. Including sub-pixels shifts in \sims galaxy images has also proven important to reduce the number of photometry outliers on real PAUCam galaxies.\\

\sims images use background cutouts from real PAUCam images. These cutouts can contain artifacts such as other galaxies, cosmic rays or crosstalk. This has proven very important for our network, as it learns how to make predictions when they are present (see e.g. $\S$\ref{sec:deblending} and $\S$\ref{sec:pausresults:singleexp}). The background light noise patterns across the CCD are narrow band dependent and changed when the PAUCam camera was modified. For this reason, the background stamps are taken from a PAUCam image observed with the same NB filter we are simulating and  we track if it was observed before or after the camera intervention (see $\S$\ref{sec:input_data}). The galaxy signal is also wavelength dependent. Consequently, we independently sample the galaxy flux from forty flux distributions, one per NB filter.\\

The simulated galaxy is convolved with the PSF as detected in the source image of the background cutout \citep[][]{psfex}.
The distribution of PSFs in the COSMOS field is also displayed in Figure \ref{fig:PAUSsims} (bottom left panel).
Using the same PSF for the galaxy and the background noise is crucial. Otherwise, the network could artificially learn that the galaxy has a different PSF than the background and use this to estimate the clean flux, which would not work on PAUCam data. Before combining the background cutout and the galaxy, we simulate photon shot noise on the galaxy. Note that other sources of additive noise, e.g. readout or electronic noise, are not required as the background cutout already includes them. This is another benefit of using real PAUCam background stamps, as simulating realistic noise is often hard and could easily lead to differences between simulations and data.\\

As simulations are constructed from real PAUS flux measurements and PAUCam background cutouts, outlier measurements in any of these two might end up represented in the \sims images. Some examples of this could be background images with spurious effects or outlier flux measurements at the catalogue level (i.e. artificially low or high fluxes).  To reduce the number of affected \sims galaxies, the flux distribution used to create them is clipped at 0 and 1000 e$^{-}$/s. This ensures that neither negative fluxes nor artificially bright examples are represented in the image simulations. Furthermore, we have also proceeded with a visual inspection of the PAUCam images. This filters out very poor observations, but cannot deal with local effects in regions of the CCD, e.g. saturated pixels, and therefore a few outliers will still leak into the \sims catalogue.\\

The methodology used to generate \sims images is very similar to that of the \textsc{Balrog} simulations \citep{Balrog1,Balrog2}. The  main similarity between \textsc{Balrog} and \sims is that both methodologies add the simulated galaxy on real survey images. In contrast, there are also differences as \textsc{Balrog} uses \textsc{Galsim} \citep{galsim} to draw the simulated galaxies, while \sims galaxies are built with \textsc{Astropy}. Also, \sims galaxies are constructed in a super-resolution grid, which increases the resolution of small objects and allows to include sub-pixel shifts from the center of the stamp.


\subsection{Comparison between PAUCam and \sims galaxies}
\label{sec:sims_comp}

Supervised machine learning algorithms require a training sample, i.e. a set of data with a known solution that is used to find the non-linear mapping from the input to the output of the network. Having a good and large training sample is a crucial part of the training and ideally, we would train \textsc{Lumos} on a sample of PAUCam images with known photometry. However, in absence of that, we are using \sims images for training. These simulations need to be representative of the testing data, and small differences between PAUS and \sims galaxies can lead to a degradation of the predictions. \\

To test the similarity between PAUCam and \sims galaxies, we have generated a controlled sample of \simsborder-PAUCam galaxy pairs. Given a PAUS galaxy, its simulated pair is constructed with a S\'{e}rsic modelling using the same profile parameters ($r_{\rm 50}, n_{\rm s}$) and the same amount of light. The simulated background  stamp is selected from a sourceless region in the same image as its real PAUCam pair.\\ 

\begin{figure*}
\includegraphics[width= 0.99\textwidth]{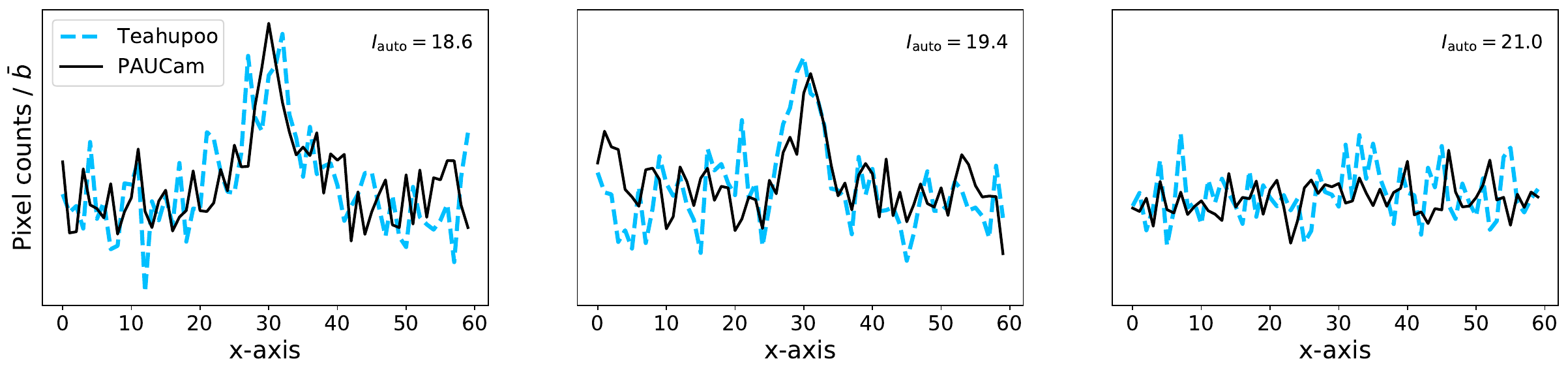}
\centering
\caption{Comparison between single pixel row light profile for pairs of PAUCam galaxies (solid black line) and \sims simulated galaxies (dashed light blue). \sims galaxies are constructed to exactly mimic its real PAUCam pair. The plot shows the pixel value along the central row of pixels (crossing the source) divided by the mean image background noise. Therefore, the central peak at x = 30 corresponds to the galaxy light contribution. From left to right, the galaxy magnitudes are $i_{\rm AB}$ = 18.6, 19.4 and 21.0.}
\label{fig:sims_comp}
\end{figure*}

Figure \ref{fig:sims_comp} shows three \sims-PAUCam galaxy pairs with  $i$-band magnitudes of 18.6, 19.4 and 21.0, respectively. The plot shows the pixel values along the central row of the image  normalised with the mean background (excluding the source) of the stamp. Therefore, pixels without galaxy light contribution should be fluctuating around unity, while  pixels with higher values will be showing the galaxy light profile. In general, PAUS and \sims galaxies fit well up to background light and shot noise fluctuations. The first plot on the left-hand side is a clear example of this. 
In the middle galaxy, the two galaxies also match reasonably well, however it also exhibits a small shift between the galaxy peaks, possibly due to an astrometry inaccuracy. The right plot shows that the comparison on fainter sources is much harder, as fainter galaxies can barely be distinguished from background fluctuations (see also Fig. \ref{fig:sims_astropy}). \\

Several effects could bring variations  between \sims and PAUCam galaxies. For instance, inaccuracies in any step of the data reduction process as e.g. the photometric calibration, the astrometry or the PSF measurement will not be fully represented in the simulations. The PSF is a clear example, as currently a single PSF measurement is taken from each PAUCam image assuming the PSF is constant across the CCD, which could potentially yield to  discrepancies between simulations and data. Additionally, if the real galaxy cannot be modelled with a single S\'{e}rsic modelling, that would also imply a difference between the two images. These discrepancies will propagate into larger errors in the flux estimation. Nonetheless, note that inaccuracies in the calibration, the modelling or the PSF would also affect the measurement with other flux estimation methods (e.g. aperture photometry or model fitting). Furthermore, \textsc{Lumos} uses both the galaxy image and the image of the modeled profile, which allows it to provide flux uncertainties that take into account discrepancies between the galaxy and the model. It also allows detecting inaccuracies in e.g. the astrometry or the PSF, which will also be considered in the uncertainty measurement (see last paragraph in \S\ref{sec:pausresults:singleexp}).

\section{Flux estimation methods}

\label{sec:methods}
In this section, we introduce other flux estimation methodologies that we will use to compare to \textsc{Lumos} performance. Particularly, we consider a profile fitting methodology (\S\ref{sec:methods:profilefitting}), aperture photometry (\S\ref{sec:methods:aperture_phot}) and a linear weighted sum of the galaxy pixels (\S\ref{sec:methods:weights}).

\subsection{Profile fitting}
\label{sec:methods:profilefitting}
In a profile fitting approach, the background subtracted galaxy image is fitted to a theoretical galaxy model to infer the profile amplitude. Assuming that the galaxy can be modelled as $I(r) = I_{\rm e} R(r)$, where $I_{\rm e}$ is the profile amplitude and R($r_{\rm i}$) corresponds to a S\'ersic light profile at pixel i, we can fit the image to the theoretical profile with 
\begin{equation}
\chi^2 = \sum_{\rm i} \frac{(f_{\rm i} - I(r_{\rm i}))^2}{\sigma_{\rm F,i}^2},
\end{equation} where i sums over pixels, f is the background subtracted flux ($f \equiv F - B$, with F and B being the total flux and the background noise) and $I(r_{\rm i})$ is the galaxy theoretical model in pixel i. Assuming Poisson errors, the previous equation becomes
\begin{equation}
\chi^2(I_e) = \sum_{\rm i} \frac{(f_{\rm i} - I_e R(r_i))^2}{I_e R(r_i) + B},
\label{eq:alpha-fitting}
\end{equation} where B is the mean background per pixel.
The total flux is measured as the $I_{\rm e}$ minimising Equation \ref{eq:alpha-fitting}. Note that the parameter $I_{\rm e}$ appears twice in the equation, which makes the closed form not feasible. Instead, we have minimised Equation \ref{eq:alpha-fitting} with a Nealder-Mead algorithm from \textsc{SciPy} \citep{scipy}.\\

\subsection{Aperture photometry}
\label{sec:methods:aperture_phot}
Aperture photometry \citep{aperture_photometry} is widely used in a large number of surveys e.g. DES \citep{DES_phot} or Pan-STARRS \citep{Pan-STARRS},  and also in PAUS. This approach measures all the pixel contributions inside an aperture of radius R with subtraction of the background light, i.e. 


\begin{equation}
    f =  \iint_{\rm R} {\rm d}A\, \left(S(r) - \Bar{b}\right),
    \label{eq:flux_ap}
\end{equation} where $S$ is the signal per pixel, $\Bar{b}$ is the mean background per pixel, R the aperture radius and A is the aperture area.\\

In PAUS, the apertures are elliptical and their areas are set to target a fixed amount of galaxy light, in our case 62.5\% of the flux. Therefore, obtaining  the total flux requires scaling the measurement (Eq. \ref{eq:flux_ap}) by 1/0.625. Given a target percentage of light, R is estimated using a simulated galaxy profile (S\'{e}rsic index, size, ellipticity) convolved with the image PSF.
The background light is measured as the mean of the pixel values within an annulus of $R_{\rm in} = 30$ pixels, $R_{\rm out} = 45$ pixels centered at the targeted galaxy. The pixels within the ring are $\sigma$-clipped beforehand, to prevent from artifacts biasing the background measurement.

\subsection{Weighted pixel sum}
\label{sec:methods:weights}
In aperture photometry, all pixels within the aperture contribute equally to the total flux, i.e. pixels at the galaxy border with low SNR and pixels at the galaxy center contribute the same. In terms of total SNR, this is not optimal, especially for small and faint galaxies where all the signal is distributed among very few pixels. Weighting differently each pixel contribution could increase the SNR of the measurements.\\

There are different choices of weights, with some providing a higher SNR than others. Indeed note that aperture photometry is just a simple case of pixel weighting, where pixels within the aperture have a unity weight and those outside do not contribute. \\

The weighting we are interested in is that giving the most optimal unbiased linear solution. This means the unbiased estimator providing the  maximum SNR, which can be written as 
\begin{equation}
    {\rm SNR} = \frac{\sum_{\rm i} w_{\rm i} m_{\rm i}}{\sqrt{\sum_{\rm i} w_{\rm i}^2 (m_{\rm i} + b_{\rm i})}},
    \label{eq:SNR}
\end{equation}  where $b_{\rm i}$, $m_{\rm i}$ and $w_{\rm i}$ are the background mean value, the signal mean value and the optimal weight in pixel i, respectively. Maximising the SNR (Eq. \ref{eq:SNR}) as a function of the pixel weights $w_i$ leads to
\begin{equation}
    w_{\rm x} =  \frac{\sum_{\rm i} m_{\rm i}}{\sum_{\rm i} m_{\rm i}^2/(m_{\rm i}+b_{\rm i})} \frac{1}{1+b_{\rm x}/m_{\rm x}},
    \label{eq:optweights}
\end{equation} where $w_{\rm x}$ is the optimal weight of pixel x (see Appendix \ref{App:flux_methods} for a more detailed derivation). This is the linear estimator providing the most precise unbiased flux measurements. However, obtaining the optimal weights in Equation \ref{eq:optweights} requires a perfect knowledge of the galaxy light profile and consequently, uncertainties in the pixel signal and background noise degrade the precision of the flux measurements. Nonetheless, this methodology puts a limit on how well linear methods can measure galaxy fluxes, which can be used to benchmark  the \textsc{Lumos} performance.

\section{\textsc{Lumos}: Measuring fluxes with a CNN}
\label{sec:lumos_general}
In this section, we describe \textsc{Lumos}, our deep learning algorithm to measure the  photometry of astronomical objects. A short introduction to the deep learning concepts needed to understand this section can be found in Appendix \ref{App:CNNs}. We will discuss \textsc{Lumos} input data (\S\ref{sec:input_data}), its architecture (\S\ref{sec:lumos}) and the training procedure (\S\ref{sec:transfer_learning}, \S\ref{sec:training_proc}).
 In sections \S\ref{sec:lumos}, \S\ref{sec:transfer_learning} and \S\ref{sec:training_proc} we describe technical details of the the network's architecture and the training procedure, which are not absolutely required to understand the main results of this work.

\subsection{Input data}
\label{sec:input_data}
The \textsc{Lumos} input consists of two types of data. The most important input includes two images of 60x60 pixels. The first one is made of an image cutout centered at the target galaxy and, given that PAUS pixel scale is 0.263, it covers 16x16 arcsec of the night sky. Although most PAUS galaxies have a half light radius between 1 and 3 pixels, which would not require such a large cutout, in  \cite{bkgnet} we already showed that the network needs larger stamps to accurately model the background noise fluctuations and scattered light patterns. The second image contains the convolved galaxy profile drawn using the parameters from an external catalogue. We have tested other possibilities like, e.g. using the true galaxy profile, the PSF profile or both separately, obtaining the best results with the convolved galaxy profile. Note that in the training phase the input galaxy cutouts are \sims image simulations (see \S\ref{sec:sims_astropy}), while in the testing phase these of real galaxy cutouts .\\

While the network can directly estimate the
photometry using only images, we have found that
additional information improves the results.
This information is the second type of input and it currently includes:
\begin{enumerate}
    \item The i-band magnitude of the target galaxy obtained from the external catalogue. This is not strictly needed as the network works without this information. However it helps providing better photometry uncertainties and so far \textsc{Lumos} requires other information (galaxy profile, coordinates) from an external catalogue anyway.
    \item The CCD coordinates. PAUCam images contain scattered light with a band dependent spatial pattern across the CCD. This makes the CCD position and the band relevant information for the network (see Figures 1 and 2 in \cite{bkgnet}). 
    \item The narrow band filter identification. The galaxy flux distribution is different for each narrow band filter. Furthermore, the scattered light pattern also depends on the NB filter (see Figure 2 in \cite{bkgnet}). Therefore, the band provides valuable extra knowledge of the expected flux and background noise pattern.
    \item A camera intervention flag. The camera was modified while observing the COSMOS field (see \S\ref{sec:data}). Therefore, the network also benefits from knowing if an image was taken before or after the camera intervention. This information is combined with that of the NB filter and given as an 80x10 trainable matrix (see \S\ref{sec:lumos} for more details). In practice, the network effectively works for all intents and purposes as having 80 different filters instead of 40.
\end{enumerate}
How these inputs are combined and given to the network is described in the next subsection.

\subsection{\textsc{Lumos} architecture}
\label{sec:lumos}

\begin{figure*}
\includegraphics[width=0.99\textwidth]{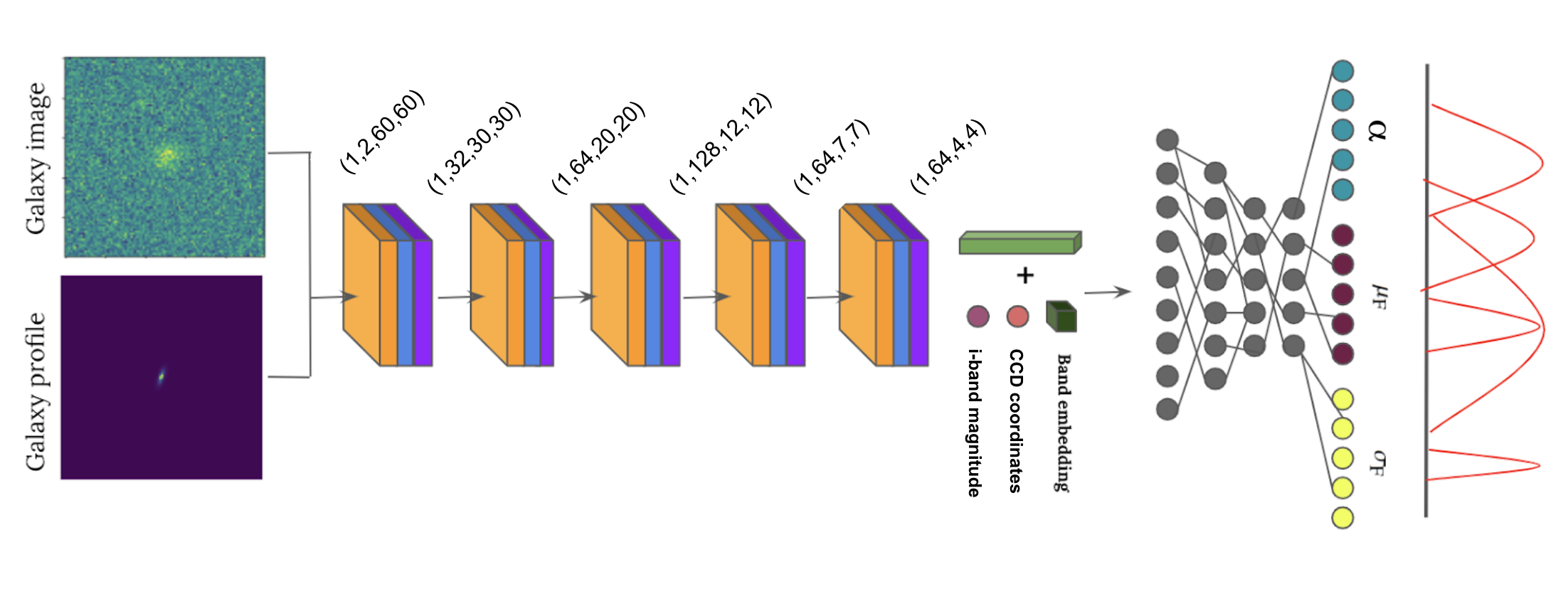}
\centering
\caption{\textsc{Lumos} architecture. Orange cubes correspond to convolutional layers, blue cubes are pooling layers and the purple ones are batch normalisation layers. The vectors between two CNN blocks corresponds to the dimension of the previous block's output and the input dimension of the following. The CNN's output is linearised (green stick), combined with external information (coordinates, NB filter and i-band magnitude) and input to a MDN. The MDN outputs a probability distribution of the total flux as a linear combination of five Gaussians. }
\label{fig:lumos}
\end{figure*}

The \textsc{Lumos} architecture (see Fig. \ref{fig:lumos}) has two differentiated parts, a Convolutional Neural Network (CNN) and a Mixture Density Network (MDN). The CNN works directly on the input images and it is built up with five blocks of convolution-pooling-batch normalisation (see Appendix \ref{App:CNNs} for a more detailed explanation). These layers are represented in Figure \ref{fig:lumos} as orange, blue and purple stacked blocks, respectively. 
The CNN's output is transformed into a 1D array including the galaxy image information and then combined with external information regarding its i-band magnitude, its position in the CCD, the NB filter it was observed with and the camera intervention flag (see \S\ref{sec:input_data}). For the band information, the network uses a 80x10 matrix, where each combination of band $\times$ camera intervention flag (before/after) is represented by 10 features to be trained.\\

The CNN's output array combined with the galaxy information is the input of the MDN. MDNs \citep{MDN} are a variant of neural networks that combine a feed-forward network and a mixture model.  The MDN returns the probability distribution of the total flux prediction as a linear combination of k distributions. These distributions could be any sort of basis function, e.g. Gaussians or multiquadratic functions. However, in this work we have only considered Gaussian distributions, in such a way that
\begin{equation}
    p(f|G) = \sum_{\rm i}^{\rm k} \alpha_{\rm i} N_{\rm i}(\mu_{\rm i},\sigma_{\rm i})\, ,
    \label{eq:MDN_prob}
\end{equation}  where $f$ is the galaxy flux, G is the galaxy data, $N_{\rm i}(\mu_{\rm i}, \sigma_{\rm i})$ is the i-th Gaussian outcome distribution and the parameters $\alpha_{i}$ are the so-called mixing coefficients. These parameters give the relative contribution of each Gaussian component to the total probability distribution.
Note that the set of mixing coefficients must sum to unity and therefore they can be understood as the prior probability of each Gaussian component. \\

The MDN in \textsc{Lumos} consists in 4 dense layers with parameters 5000:1000:100:15. It outputs five mixing coefficients ($\alpha$) together with five pairs of $(\mu, \sigma)$ parametrising the Gaussian components (N). This kind of network architecture was already used on PAUS data in \cite{Deepz}, where it was used to predict redshift probability distributions. \\

The choice of loss function is a crucial step in the construction of a neural network. \textsc{Lumos} combines two loss functions, the first as in Equation \ref{eq:lossGaussian_1} and a modified version of it in Equation \ref{eq:lossGaussian2}.\\

A common loss function for a Gaussian MDN is 
\begin{equation}
{\rm \mathcal{L}_{\rm MDN}} = \sum_{\rm i=1}^{k} \left [  \log(\alpha_{\rm i}) - \frac{(f_{\rm i} - \bar{f})^2}{\sigma_{\rm i}^2} - 2\log{(\sigma_{\rm i})}\right] ,
\label{eq:lossGaussian_1}
\end{equation}  which corresponds to the negative log likelihood of a Gaussian distribution with mean $\bar{f}$ and variance $\sigma_{\rm i}^2$.\\

In \textsc{Lumos}, we have also considered a variation of the previous loss motivated by the fact that physical galaxies have positive flux values. This alternative loss also corresponds to the Gaussian negative log-likelihood, but integrated from 0 to $\infty$, which changes the PDF normalisation.\\

The logarithm of a Gaussian PDF ($G(x)$) integrated from 0 to $\infty$ is
\begin{eqnarray}
    \nonumber 
    \log\left({G(x)}\right) = \log{\left( \frac{2}{\pi} \right)} + \log{\left(\exp \left(\frac{-\frac{1}{2}(x - \mu)^2}{\sigma^2}\right) \right)} + \\ 
    \log {\left(\sigma {\rm Erf}\left(\frac{\mu}{\sqrt{2}\sigma}\right)+\sigma\right)}\: ,
\end{eqnarray} which leads to the following loss function
\begin{eqnarray}
    \nonumber 
    {\rm \mathcal{L}_{\rm MDN}} = \sum_{\rm i=1}^{k}  \log{(\alpha_{\rm i})} - \frac{(f_{\rm i} - \bar{f})^2}{\sigma_{\rm i}^2} - 2\log{(\sigma_{\rm i})} - \\
    \log{\left({\rm Erf}\left(\frac{\bar{f}}{\sqrt{2}\sigma_{\rm i}} \right)+1\right)}\: ,
    \label{eq:lossGaussian2}
\end{eqnarray} where again $\bar{f}$ is the true flux and $\sigma_{\rm i}$ is the flux uncertainty. Therefore, the truncation of the Gaussian distribution effectively corresponds to an additional term in the loss function. \textsc{Lumos} combines both loss functions and uses Equation \ref{eq:lossGaussian_1} on objects with $i_{\rm AB} <20.5$ and Equation \ref{eq:lossGaussian2} for the rest of the sample. Choosing a single loos function for all magnitudes also works, however combining the two improves the photometry for the faintest and the brightest galaxies. 

\subsection{Unsupervised transfer learning}
\label{sec:transfer_learning}

Transfer learning \citep{TL2,TL} is a deep learning technique that aims to adapt a model trained to make predictions on a particular task to work on a similar but different problem. One example is a classifier trained to distinguish between cats and dogs adapted to distinguish horses and zebras. Instead of training from scratch, the zebra-horse classifier takes the parameters of the cat-dog classifier as initial parameters, in such a way that the network has already learnt to extract shared features like e.g. detect edges or shapes. Such mutual features are commonly extracted in the shallower layers of the network, while deeper layers pick up more subtle data traits. For this reason, many times transfer learning only requires training deeper layers of the network, while shallower layers can be shared among the different networks. \\

The same idea can be applied to adapt models trained on simulations to perform well on data \citep{TL_simsdata,Deepz}. To train a supervised network one needs data with a known solution (labeled data). Many times there are labeled data available, but not enough to train a network from scratch. A possible solution in such cases is to train the network on simulations and use a small labeled dataset to adapt the model to the data. This requires two consecutive trainings, one initial on simulations and once this finishes, an additional one on data with the network parameters from the training on simulations as a starting point. \citealt{TL_helena} also explores the possibility of using transfer learning to adapt a model trained on data from one survey to another one. Particularly, it adapts a morphology classifier trained on SDSS images to work on DES images. \\

\textsc{Lumos} is trained on \sims image simulations (see \S\ref{sec:data} for more details) and we cannot apply supervised transfer learning as there are no data with known photometry. Instead, we will use the compatibility of independent observations of the same galaxy in the same NB filter to apply what we call \textit{unsupervised transfer learning}. \\

For that, we have collected a set of PAUCam observed galaxy pairs, with two independent images of the same galaxy observed with the same narrow band filter. \textsc{Lumos} should predict compatible flux PDFs for the two observations of the same object, learning to ignore differences in e.g. the background noise or the PSF. Therefore, after training \textsc{Lumos} on \sims image simulations, we retrain it  on the set of PAUCam galaxy pairs, forcing compatibility between the two flux measurements. With this procedure, we make sure that \textsc{Lumos} has seen real data before evaluating the network on the test sample. \\

The unsupervised transfer learning loss function is constructed comparing the probability distribution of two observations. Before comparing the PDFs, these need to be calibrated with the image zero-point (see \S\ref{sec:pausdata} for more details). The PDFs are parametrised with five Gaussian distributions (as provided by \textsc{Lumos}, see \S\ref{sec:lumos}), in such a way that each Gaussian component in the first observation is compared to all the components in the second observation. The negative log likelihood of the difference between the two predicted PDFs takes the form of
\begin{eqnarray}
\nonumber
{\rm \mathcal{L}_{\rm UTL}} = \sum_{\rm i} \sum_{\rm j} \log{\alpha_{\rm i}} + \log{\beta_{\rm j}} - \frac{1}{2}\frac{(f_{i} -f_{j})^2}{\sigma_{\rm i}^2 + \sigma_{\rm j}^2}  \\ \nonumber  
-\sqrt{\sigma_{f_{\rm i}}^2 + \sigma_{f_{\rm j}}^2} - 
\log{\left[1 +{\rm Erf}\left(\frac{f_{\rm i}}{\sqrt{2} \sigma_{\rm i}} \right)\right] }  \\ \nonumber
- \log{\left[1 +{\rm Erf}\left(\frac{f_{\rm j }}{\sqrt{2} \sigma_{\rm j}} \right)\right]}  \\ 
+\log{\left[1 + {\rm Erf}\left(\frac{f_{\rm i} / \sigma_{\rm i}^2 + f_{\rm j} / \sigma_{\rm j}^2}{\sqrt{1 /\sigma_{\rm i}^2 + 1/\sigma_{\rm j}^2}} \right)\right]},
\label{eq:unsupervised_loss}
\end{eqnarray} where $f_{\rm i}$ ($f_{\rm j}$) is the i-th (j-th) Gaussian component of the first (second) exposure, and similarly for $\sigma$.\\


\subsection{Training procedure}
\label{sec:training_proc}
 \textsc{Lumos} is trained with 500,000 simulated images constructed combining simulated galaxies and real PAUCam backgrounds (see \S\ref{sec:sims_astropy}). The sample is split in 90\% for training and 10\% for validation. We also generate 10,000 independent simulated galaxies for testing. \textsc{Lumos} is trained for 100 epochs with an initial learning rate of $10^{-4}$, which is reduced a factor of 10 every 40 epochs. We use \textsc{Adam} \citep{Adam} as optimisation algorithm. The training
 takes about 20 hours using an NVIDIA TITAN V GPU. \\

The network is trained with a batch size of 500 galaxies, which is a relatively large batch size. As simulations are constructed from real PAUS flux measurements and PAUCam background cutouts, \sims image simulations might contain some outliers. We have applied a filtering to reduce them (see the last paragraph in \S\ref{sec:sims_astropy}), however a few poor examples can still be part of \sims images and these are very difficult to detect. Having a large batch size reduces their effect on the overall loss function and consequently in the training.\\

After the supervised training, we apply the unsupervised transfer learning. This part is trained with 20,000 galaxy pairs. A danger of applying unsupervised transfer learning is that nothing prevents the network from biasing the flux predictions, since Equation \ref{eq:unsupervised_loss} does not directly constrain the flux prediction but the pairwise consistency. For this reason, we have also included a supervised training with 20,000 simulations. The loss function ends up combining Equations \ref{eq:lossGaussian_1}, \ref{eq:lossGaussian2} and \ref{eq:unsupervised_loss} and therefore becomes
\begin{equation}
    {\rm \mathcal{L}} = {\rm \mathcal{L}_{\rm MDN}} + {\rm \mathcal{L}_{\rm UTL}}. 
\end{equation} These two losses are weighted equally in the loss function. Nevertheless, this is not strictly necessary and one could consider weighting them differently.\\

\textsc{Lumos} has already been trained on reliable simulations, therefore its parameters should be close to optimal before the unsupervised transfer learning training. Consequently, instead of training all the network parameters, the transfer learning only varies the parameters in the last linear layer, while all the others are frozen. The unsupervised transfer learning phase is trained for 100 epochs, with an initial learning rate of $10^{-5}$, which is reduced by a factor of 10 every 50 epochs. 


\section{\textsc{Lumos} flux measurements on  simulations}
\label{sec:resultssims}
In this section we will test \textsc{Lumos} on \sims galaxies. We will first validate the flux probability distributions (\S\ref{sec:resultssims:pit}), followed by a comparison with other flux estimation methods (\S\ref{sec:meth_comp}). Finally we will test how well \textsc{Lumos} performs on blended galaxies (\S\ref{sec:deblending}).

\subsection{Flux probability distributions}
\label{sec:resultssims:pit}
\begin{figure}
\includegraphics[width=0.45\textwidth]{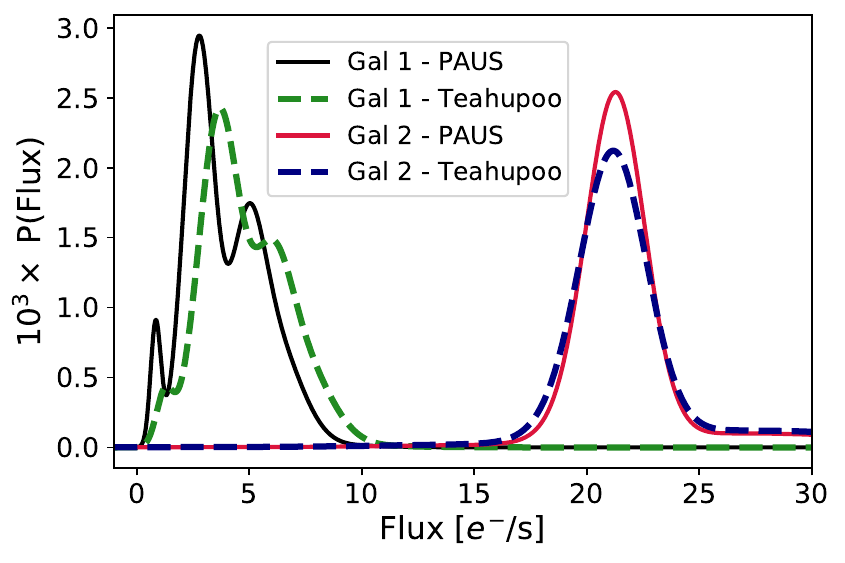}
\centering
\caption{Flux probability distributions provided by Lumos for two PAUCam galaxies (Gal1, Gal2, solid lines) and their \sims imitations (dashed lines).}
\label{fig:ex_pdf}
\end{figure}

Most photometry algorithms only provide a flux measurement and its uncertainty. In contrast, \textsc{Lumos} provides the flux probability distributions as a linear combination of five  Gaussians (see \S\ref{sec:lumos}).  Figure \ref{fig:ex_pdf} shows the predicted flux PDFs for two PAUCam  galaxies (solid lines) and two \sims galaxies mimicking them (dashed lines) (see \S\ref{sec:sims_comp}). The flux probability distributions on data and on the simulations are very similar, providing additional confidence on the reliability of \sims galaxies.\\ 

Also, notice that the resulting PDFs are not Gaussian. This can especially be seen in the faintest galaxy, which displays secondary peaks on the left and right of the main one. This type of PDF is common in \textsc{Lumos} predictions, where fainter galaxies exhibit more non-gaussianities than brighter ones. In general, \textsc{Lumos} PDFs are more Gaussian at redder bands, where galaxies are also brighter. At the blue end, many PAUS galaxies have fluxes very close to 0 for which \textsc{Lumos} commonly provides very non-Gaussian PDFs (see \S\ref{sec:pausresults:coadds} for further discussion). \\

\subsubsection{Probability Integral Transform (PIT) on simulations}
The Probability Integral Transform (PIT) \citep{PIT1,PIT2,PIT3} tests the quality of the probability distribution and it is defined by 
\begin{equation}
    {\rm PIT} \equiv \int_{\rm -\infty}^{f^{*}} {\rm d}f\,\phi(f)\, 
\end{equation}
where $f^{*}$ is the true flux value and $\phi(f)$ is the probability distribution. When $\phi(f)$ faithfully represents the true value, the PIT distribution is the uniform distribution U[0,1]. \\

\begin{figure}
\includegraphics[width=0.45\textwidth]{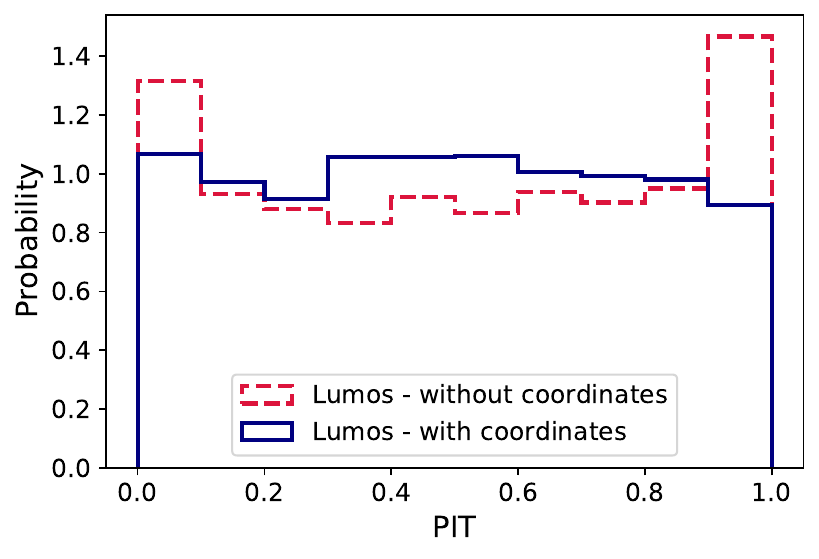}
\centering
\caption{PIT distribution of the \textsc{Lumos} flux PDFs on a set of 10,000 \sims galaxies. We have tested \textsc{Lumos} PDFs with (solid blue line) and  without (red dashed line) including the CCD coordinates. If the CCD coordinates are not included, \textsc{Lumos} provides outliers (peaks at 0 and 1) corresponding to scattered light affected objects.}
\label{fig:PIT_erf}
\end{figure}

In Figure \ref{fig:PIT_erf}, we have estimated
the PIT value for  10,000 \sims galaxies with known true flux.  The plot shows two distributions, one including (solid blue line) and another not (dashed red line) including the CCD coordinates of galaxies in the training and test set. When the training does not include the coordinates, the PIT distribution displays two peaks of outliers at the first and last bin of the plot. These outliers correspond to galaxies with strongly varying  background light that require accurate knowledge of the background noise patterns in the different NB filters. \\

In contrast, when the training includes the CCD coordinates, the PIT test displays a flat U[0,1], showing that \textsc{Lumos} provides robust flux probability distributions and that CCD coordinates are essential information for cutouts with varying backgrounds. This is consistent with the results in Figure 7 of \cite{bkgnet}, where the CCD coordinates proved essential to predict accurate backgrounds (solid black line).\\

\subsubsection{Single flux and flux uncertainty measurements}
Although the flux PDF provides more information, many applications require a single flux measurement and uncertainty. Flux point-like estimates can be computed with different statistical estimators as the mean, the median or the peak. For (almost) Gaussian PDFs, these estimators provide very similar flux measurements. However when the PDFs move away from gaussianity, these estimators can provide significant differences among them. As an example, in \textsc{Lumos} multiple peaked distributions tend to provide higher flux measurements with the median than with the peak. This is because these PDFs commonly belong to faint objects with the main peak very close to zero and secondary peaks and the tail moving towards higher fluxes (see Fig. \ref{fig:coadd_pdf} for an example).\\

For the flux uncertainty, the most straightforward estimator  is the standard deviation. However, another possibility is 
\begin{equation}
    \sigma_{\rm 68} \equiv \frac{1}{2}[q_{\rm 84} - q_{\rm 16}]\, ,
    \label{eq:sigma68}
\end{equation} where $q_{\rm 16}$ ($q_{\rm 84}$) are the 16-th (84-th) quantiles. 
For Gaussian distributions, these two estimators coincide. However,  in the case of non-Gaussian PDFs, $\sigma_{\rm 68}$ is more robust towards distributions with tails but provides higher uncertainties in the presence of multiple peaks (see \S\ref{sec:pausresults:coadds} for further discussion).\\

Even though the median or $\sigma_{\rm 68}$ are more robust with noisy PDFs, they require the explicit PDF construction from the Gaussian parametrisation provided by \textsc{Lumos}. This is time consuming, since we already have more than 10 million galaxy exposures in the small COSMOS field. A fast alternative is to analytically determine the mean and the variance from the Gaussian component parameters. The mean flux ($f$) is estimated as
\begin{equation}
    f = \sum_{\rm i} \alpha_{\rm i} · \mu_{\rm i}\, , 
    \label{eq:MDN_mean}
\end{equation} where $\alpha_{\rm i}$ and $\mu_{\rm i}$ are the mixing coefficient and the expected value of the i-th Gaussian component, respectively. The associated variance ($\sigma_{\rm f}^2$) is then given by

\begin{equation}
    \sigma^2_{\rm f} = \sum_{\rm i} \left[\alpha_{\rm i}\  \left(\sigma_{\rm i}^2 +\ \norm{\mu_{\rm i} - \sum_{\rm j}\alpha_{\rm j} \mu_{\rm j})}^2\right)     \right], 
    \label{eq:MDN_var}
\end{equation} where  $\sigma_{\rm i}^2$ is the variance of the 
$i$-th Gaussian component.\\


\subsection{Comparison with different flux estimation methods}
\label{sec:meth_comp}
While \textsc{Lumos} has proven to provide reliable flux probability distributions, other methodologies as model fitting or  aperture photometry are also able to provide accurate flux estimates \citep[e.g.][]{tracotr, DES_phot, kids_phot}.
In this section, we will use simulations to compare the performance of \textsc{Lumos}  with a profile fitting method (\S\ref{sec:methods:profilefitting}), aperture photometry (\S\ref{sec:methods:aperture_phot}) and a linear weighted sum of pixels (\S\ref{sec:methods:weights}).  To quantify the quality of the flux measurements, we will use
\begin{align}
    &{\rm \textbf{Bias}:} &{\rm Median}\left[(f - f_{\rm 0}) / f_{\rm 0}\right],\\
    &{\rm \textbf{Dispersion:}} &\sigma_{68}\left[(f - f_{\rm 0})/f_{\rm 0}\right],
\end{align} where $f_{\rm 0}$ is the ground truth flux.\\

\begin{figure*}
\includegraphics[width=0.89\textwidth]{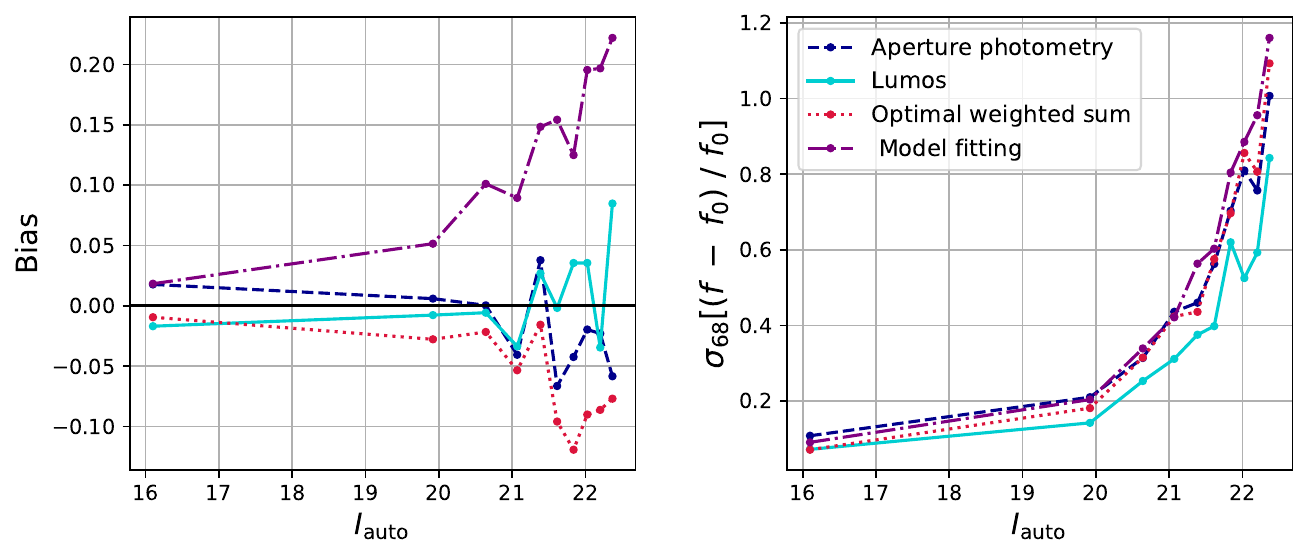}
\centering
\caption{Comparison of the bias (left panel) and the dispersion (right panel) among the flux measurements with \textsc{Lumos}, aperture photometry, model fitting and optimal pixel weighting. These results include galaxy image simulations in the 40 PAUS NB filters. The I-band magnitude corresponds to the AUTO magnitude as measured by the HST-ACS on the COSMOS field.}
\label{fig:sims_comparison_algorithms}
\end{figure*}

Figure \ref{fig:sims_comparison_algorithms} compares the bias (left panel) and the dispersion (right panel) in the flux predictions as a function of the $I_{\rm auto}$ magnitude for the four methods. The model fitting method (purple dashed dotted line) displays a systematic increment of the bias with magnitude, with a 20\% bias at the faint end. PAUS galaxies are already hard to distinguish from background fluctuations for magnitudes $i_{\rm AB} > 20$ (see Figure \ref{fig:PAUSsims}), which could be severely complicate the fitting at fainter magnitudes. \\

The second method is the pixel weighted sum. It is unbiased for objects with $i < 21$, but fainter objects are 10\% biased. While the optimal weights (Eq. \ref{eq:optweights}) ensure unbiased flux estimates, these also require perfect knowledge of the galaxy profile and the background light. On simulations, the galaxy light distribution is known, however the background light is not. Therefore, at the faint end where the galaxy signal is comparable to the background fluctuations, the weights (Eq. \ref{eq:optweights}) seem to be very sensitive to inaccuracies in the background noise. In contrast,  aperture photometry and \textsc{Lumos} provide unbiased estimates  to a 5\% level up to magnitude 22.\\

In terms of dispersion, \textsc{Lumos} is the most precise method with  $\sigma_{\rm 68} = 0.36$. This implies a 28\% improvement with respect to the linear pixel weighting method, which is the second best method with  $\sigma_{\rm 68} = 0.47$. As expected, the optimal weighted sum is the most precise unbiased linear method, but it degrades at fainter magnitudes. \textsc{Lumos} overcomes the linear optimal weighting with a non-linear mapping and provides good flux estimates at all magnitudes.\\

The previous results combine measurements from all NB filters. Considering measurements from each NB filter independently, bluer bands show a higher dispersion, which is expected since their SNR is lower. Furthermore, the flux measurements are unbiased to a 3\% level in all NB filters and these also show unbiased for all galaxy sizes ($r_{\rm 50}$ and PSFs.

\subsection{Deblending with \textsc{Lumos}}
\label{sec:deblending}
Blending is the superposition of galaxies with other astrophysical objects along the line of sight. It affects the photometric and shape measurements contributing to systematics in weak lensing studies \citep{deblending_weak_lensing}. Deblending will be a challenge for future ground-based photometric surveys such as Rubin \citep{LSST} or {\it Euclid} \citep{Euclid} and it has recently been approached with deep learning techniques \citep[e.g.][]{deblending_weak_lensing,deblending1}.\\

In this section, we test if \textsc{Lumos} can extract the galaxy photometry when the target galaxy is blended with another object. Even though \textsc{Lumos} is not explicitly constructed to predict the photometry in the presence of other galaxies, the simulated training sample is built using background cutouts from PAUCam images (see \S\ref{sec:data}). These image cutouts are centered at random CCD positions, without minding if there are any other artifacts nearby.
Consequently, the training sample contains examples of blended galaxies to learn from. This is one benefit of machine learning algorithms. These are flexible enough to learn how to extract the photometry of blended sources by only including examples in the training sample, without explicitly constructing the algorithm for this task.\\

We have generated 3600 60x60 pixel realisations of the same target \sims galaxy, which is placed at the central pixel of the stamp. Each of these realisations also contains the same PAUS galaxy centered on a different pixel at a time. The PAUS galaxy moves across the stamp in steps of one pixel, in such a way that the realisation where it is located at the central pixel corresponds to a total blending with the \sims galaxy. At the end of the day, with all the realisation the PAUS galaxy covers all the pixels in the stamp.\\

\begin{figure}
\includegraphics[width=0.45\textwidth]{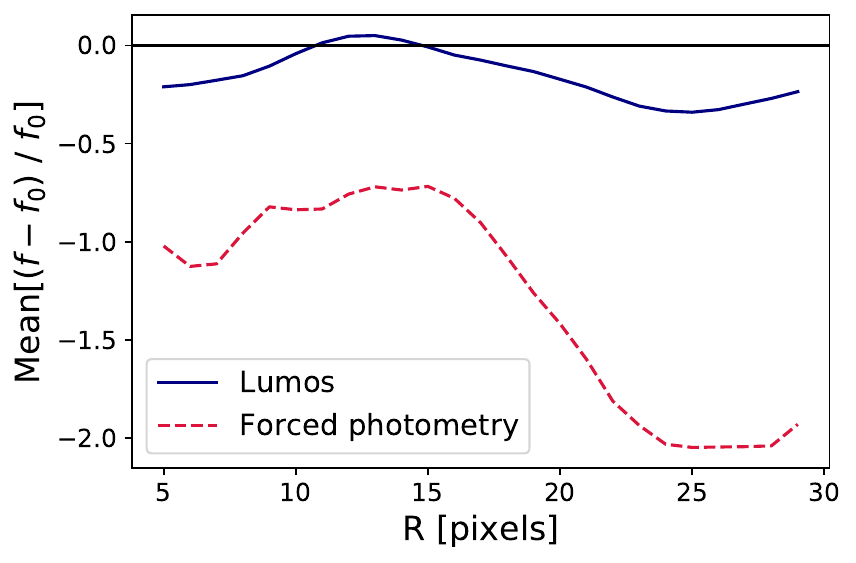}
\centering
\caption{Accuracy in the flux measurement when the target galaxy $(i_{\rm AB = 22}$) is blended with another source $(i_{\rm AB = 20}$) as a function of the relative distance between \sims target galaxy and the 'blending' PAUS galaxy (R). The solid blue line corresponds to \textsc{Lumos}, while the  red dashed line is forced photometry.}
\label{fig:deblending}
\end{figure}

Figure \ref{fig:deblending} shows the accuracy in the flux measurement as a function of distance to the overlapping source. The target galaxy is fainter than the overlapping one, with $i_{\rm AB} = 22$ and $i_{\rm AB} = 20$, respectively. With aperture photometry (dashed red line), the flux is considerably biased for all distances R. This is expected since summing all contributions within the aperture does not differentiate if the light belongs to the target source or the overlapping one. Therefore, for $R < 15$, the bias measurement is caused by light from the overlapping source accounted inside the aperture. At larger values of R, the background noise prediction is also affected by the overlapping source. In the PAUS aperture photometry pipeline, the background is estimated within a 15 pixel wide annulus, located at 30 pixels from the target source. When the overlapping source is very bright, as it happens in this example, it can affect the background prediction and the flux prediction at the same time, which degrades the performance even more.\\

On the other hand, \textsc{Lumos} (blue solid line) extracts better the photometry of blended galaxies. The relative bias of the measurement in Figure \ref{fig:deblending} fluctuates around 2-10\% for different distances R. Unlike aperture photometry, \textsc{Lumos} is able to distinguish between the two galaxies and consider the overlapping one a source of noise. \\

\begin{figure}
\includegraphics[width=0.45\textwidth]{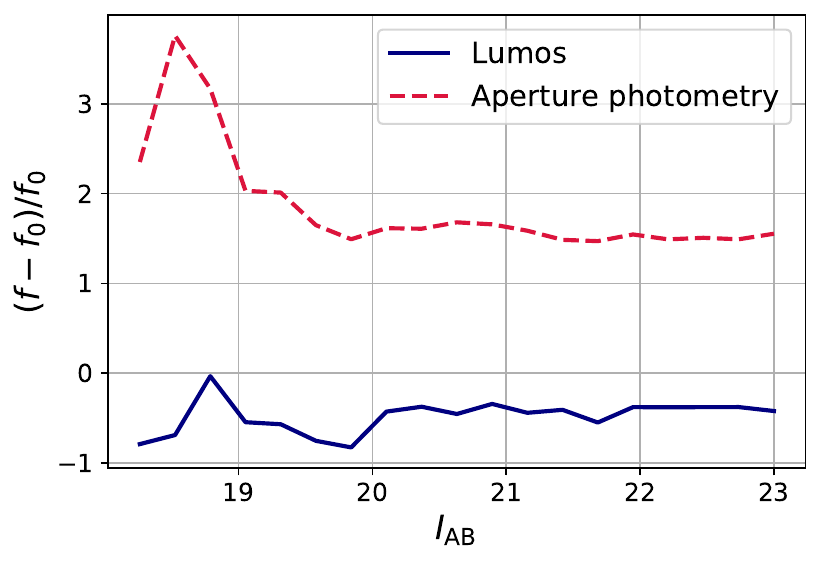}
\includegraphics[width=0.45\textwidth]{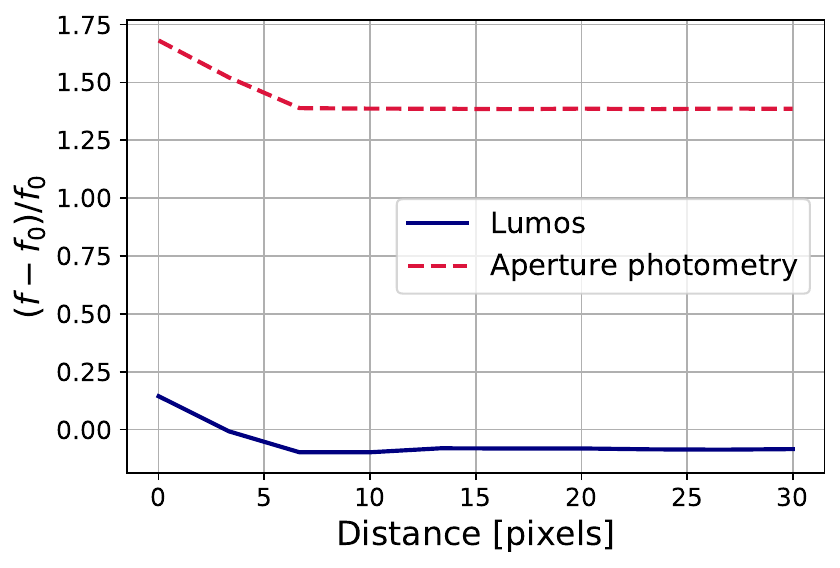}
\centering
\caption{Accuracy in the flux predictions in the presence of overlapping sources for \textsc{Lumos} (solid blue line) and aperture photometry (dashed red line) as a function of \emph{Top}: magnitude of the overlapping source. \emph{Bottom} Distance in pixels between the target and the overlapping source.}
\label{fig:deblending2}
\end{figure}

In Figure \ref{fig:deblending2}, we have explored more the \textsc{Lumos} deblending capability as a function of the magnitude and distance to the overlapping galaxy. For the top plot, we have simulated 20 image cutouts with an $i_{\rm AB} \approx 21$ galaxy at the center (the same for the twenty realisations). In each of the cutouts, we have included a second source always located at five pixels from the center, but which varies brightness among realisations. The plot shows the photometry accuracy with \textsc{Lumos} (blue solid line) and aperture photometry (red dashed line) for different magnitudes of the overlapping sources. For this study, we have also applied a $\sigma$-clipping of the annulus so that the background measurement in the aperture photometry  is more robust. As indicated in the previous test, \textsc{Lumos} shows more robust towards overlapping nearby sources, even when these are bright. In all cases, aperture photometry provides biased measurements due to the proximity of the source, which is always accounted within the aperture pixels.\\

For the bottom plot, we have proceeded similarly. We have generated ten realisations of the same galaxy with magnitude $i_{\rm AB} \approx 21$ and in each of the cutouts, we have included a second source with $i_{\rm AB} \approx 22$, but located at a different distance from the target source. The plot exhibits the photometry accuracy as a function of the distance between the target and the overlapping source. Again, the plot shows much better accuracy for \textsc{Lumos} than aperture photometry.

\section{\textsc{Lumos} photometry on PAUS data}
\label{sec:pausresults}
In this section we present the photometry extracted from PAUCam images in the COSMOS field with \textsc{Lumos}. First, we show single observation measurements (\S\ref{sec:pausresults:singleexp}) and compare them to SDSS measurements (\S\ref{sec:pausresults:sdss}). We then discuss the co-added flux measurements (\S\ref{sec:pausresults:coadds}) and show the photometric redshift results with \textsc{Lumos} photometry (\S\ref{sec:photoz}).

\subsection{Single exposure measurements}
\label{sec:pausresults:singleexp}
PAUS has taken around 10,000 images in the COSMOS field (\S\ref{sec:pausdata}), which contain ten million galaxy observations. In this section we will only show the results on the spectroscopic sample with $i_{\rm AB}<22.5$, which are $\approx$ 3 million exposures from 15,000 galaxies. For the targeting, we use the  \cite{Ilbert2008cat} catalogue, the same external catalogue that \textsc{MEMBA} (the PAUS aperture photometry algorithm, see \S\ref{sec:pausdata}) uses.
Given a PAUCam image, \textsc{Lumos} matches it with the detection catalogue and creates the cutouts around the sources. Each source is also matched to the external catalogue, which provides a value for the half light radius, the ellipticity and the S\'{e}rsic index. From these values, \textsc{Lumos} generates the modelled galaxy profiles (see \S\ref{sec:input_data}). Then, it provides the flux probability distribution of the observed galaxies.\\

\begin{figure}
\includegraphics[width=0.45\textwidth]{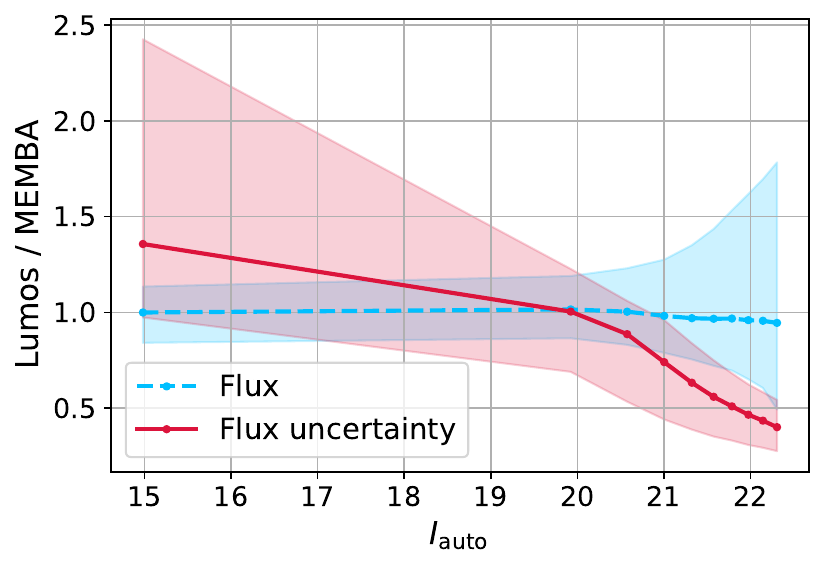}
\centering
\caption{Flux and flux uncertainty ratios between \textsc{Lumos} and \textsc{MEMBA} photometry in equally populated magnitude bins. The shaded areas correspond to the 16th and 84th quantiles.}
\label{fig:ratios}
\end{figure}

\subsubsection{Flux and flux error measurements}
In this section, we will use flux and flux error point estimates calculated as the mean and the variance of the \textsc{Lumos} flux PDFs (see \S\ref{sec:resultssims:pit}), which corresponds to the total galaxy flux. For \textsc{MEMBA}, we will use flux measurements from aperture photometry using the background subtraction from \textsc{BKGnet}, which has proven more accurate than that estimated with an annulus. Figure \ref{fig:ratios} shows the flux (dashed blue line) and flux error (solid red line) ratios between \textsc{Lumos} and \textsc{MEMBA} photometry in all NB filters. The shaded areas correspond to the 16th and 84th quantiles. For the full sample, the flux ratio between the two photometries is 0.99. In magnitude bins, this ratio oscillates between 0.95 and 1.02, with \textsc{Lumos} measuring $\approx$ 4\% less flux in the brightest ($i_{\rm AB} < 18$) and the faintest ($i_{\rm AB} >22$) bins. At the faintest end, the spread in the flux ratio increases, which is natural since these galaxies are noisier. Studying each NB filter independently, all the ratios but those from the three bluest bands ('NB455', 'NB465' and 'NB475') oscillate between 0.95 and 1.03. The three bluest bands display a $\approx0.9$ ratio between \textsc{MEMBA} and \textsc{Lumos}, which is attributed to very faint galaxies with negative flux measurements in \textsc{MEMBA}, which are not allowed in \textsc{Lumos}. \textsc{MEMBA} has proven accurate enough to obtain very precise photo-zs, therefore measuring similar fluxes with \textsc{MEMBA} and \textsc{Lumos} is a good first test.\\

Altogether, \textsc{Lumos} also provides 40\% lower flux uncertainties than \textsc{MEMBA} displaying a lower error for 85\% of the measurements. The ratio between \textsc{Lumos} and \textsc{MEMBA} flux uncertainties (Fig \ref{fig:ratios}) is not constant with magnitude. \textsc{Lumos} shows 60\% lower errors for objects with $i_{\rm AB}>22$. This number monotonically decreases to  e.g. 30\% at $i_{\rm AB}=21$ or 10\% at  $i_{\rm AB}=20.5$, while some of the brightest objects display lower errors with \textsc{MEMBA}. At the brightest end, one can note a large scatter of the error ratio. This is attributed to some very bright galaxies with a significantly large error with \textsc{Lumos}. However, note that aperture photometry provides a purely statistical error, while in \textsc{Lumos} any additional source of uncertainty, e.g. an artifact in the cutout, inaccuracies in the profile parameters used to infer the photometry ($n_{\rm s}$, $r_{\rm 50}$) or  data reduction issues underrepresented in the training simulations will be also accounted in the error estimate. As \textsc{Lumos} is provided with both the galaxy image and the model, it is able to capture discrepancies or potential sources of inaccuracies and account for them in the flux uncertainty. Following this line of study, inaccuracies in the profile parameters would most likely have a larger impact on the photometry of large, bright and resolved galaxies, where e.g. slightly underestimating $r_{50}$ could easily lead to a quite biased flux measurement. This could also explain the increment of spread in the uncertainty ratio at the brightest end. \\

\subsubsection{Colour histograms}

Assuming that galaxies have an underlying distribution of colours,
the width of the colour histograms is an estimation of the uncertainty in the photometry measurements. The intrinsic width of the colour histogram is broaden by photometry uncertainties and consequently, the photometry providing narrower colour histograms is that with lowest uncertainties. Using colour histograms to compare photometries was used in \cite{GAMA_angus}, where they presented and applied \textsc{LAMBDAR} to improve the Galaxy and Mass Assembly  \citep[GAMA,][]{GAMA} photometry .\\

Figure \ref{fig:colorhist} shows the NB785-NB795 colour distribution (more colour histograms can be found in Appendix \ref{App:colorhists}). By eye it can be already noted that \textsc{Lumos} provides a narrower colour distribution than \textsc{MEMBA}. We have estimated the width of such colour histograms with $\sigma_{\rm 68}$ (Eq. \ref{eq:sigma68}) and $\sigma_{\rm 95}$ (equivalent to $\sigma_{\rm 68}$ but considering the 2.5 and 97.5 quantiles, i.e. the width accounts for 95\% of the data). Concretely, \textsc{MEMBA} provides $\sigma_{\rm 68}$ = 0.26, while \textsc{Lumos} results in  $\sigma_{\rm 68}$ = 0.19, which corresponds to a 30\% lower effective width. Considering $\sigma_{\rm 95}$, \textsc{Lumos} reduces the width a factor of $\approx$3, from 0.74 to 0.41. This also suggests that \textsc{Lumos} reduced the number of photometry outliers, which are not affecting $\sigma_{\rm 68}$ but enlarge $\sigma_{\rm 95}$. Such photometric outliers are located asymmetrically on the tails of the distribution, which triggers the skewness of the histograms and therefore a shift in the median of the \textsc{MEMBA} histogram with respect to that of \textsc{Lumos}. This can be noted in the NB785-NB795 colour histogram, but also in other colour histograms in Figure \ref{fig:color_hists}. \\

All the other narrow bands also show narrower color histograms with \textsc{Lumos} (see Appendix \ref{App:colorhists} for more details). Furthermore, the relative difference in $\sigma_{\rm 95}$ is systematically higher than with $\sigma_{\rm 68}$. This is likely related with exposures with noisy photometry and outliers, which lay in the tails of the colour histograms (see Appendix \ref{App:colorhists} for more details).

\begin{figure}
\includegraphics[width=0.45\textwidth]{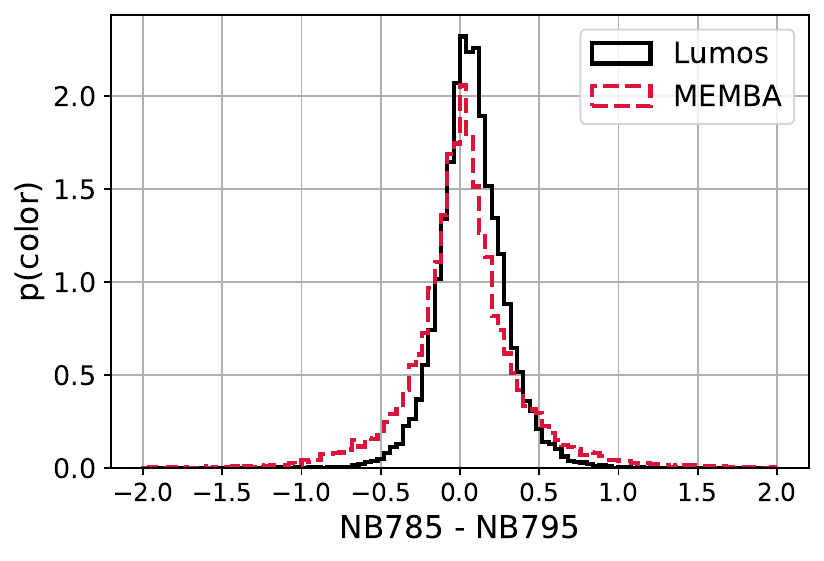}
\centering
\caption{\textsc{MEMBA} (red dashed line) and \textsc{Lumos} (solid black line) colour histogram  for NB colour (NB785-NB795). Note that the photometry with lower uncertainties is that displaying a narrower colour histogram, which in this case is the \textsc{Lumos} photometry}.
\label{fig:colorhist}
\end{figure}

\subsubsection{Validation of the flux uncertainties}
To test \textsc{Lumos} flux uncertainties and ensure that these are not artificially low, we have made use of PAUS taking multiple observations of the same galaxy in the same narrow band filter.  Given two observations of the same galaxy in the same NB, their flux measurements and uncertainties must be compatible. This is formulated as
\begin{equation}
    D \equiv \frac{(f_{\rm 1} - f_{\rm 2})}{\sqrt{(\sigma^2_{\rm 1} + \sigma^2_{\rm 2})}}\, ,
    \label{eq:Duplicates}
\end{equation} where $f_{\rm 1}$, $f_{\rm 2}$ are the flux estimates of two exposures of the same object and $\sigma_{\rm 1}$, $\sigma_{\rm 2}$ are their associated uncertainties. \\

\begin{figure}
\includegraphics[width=0.47\textwidth]{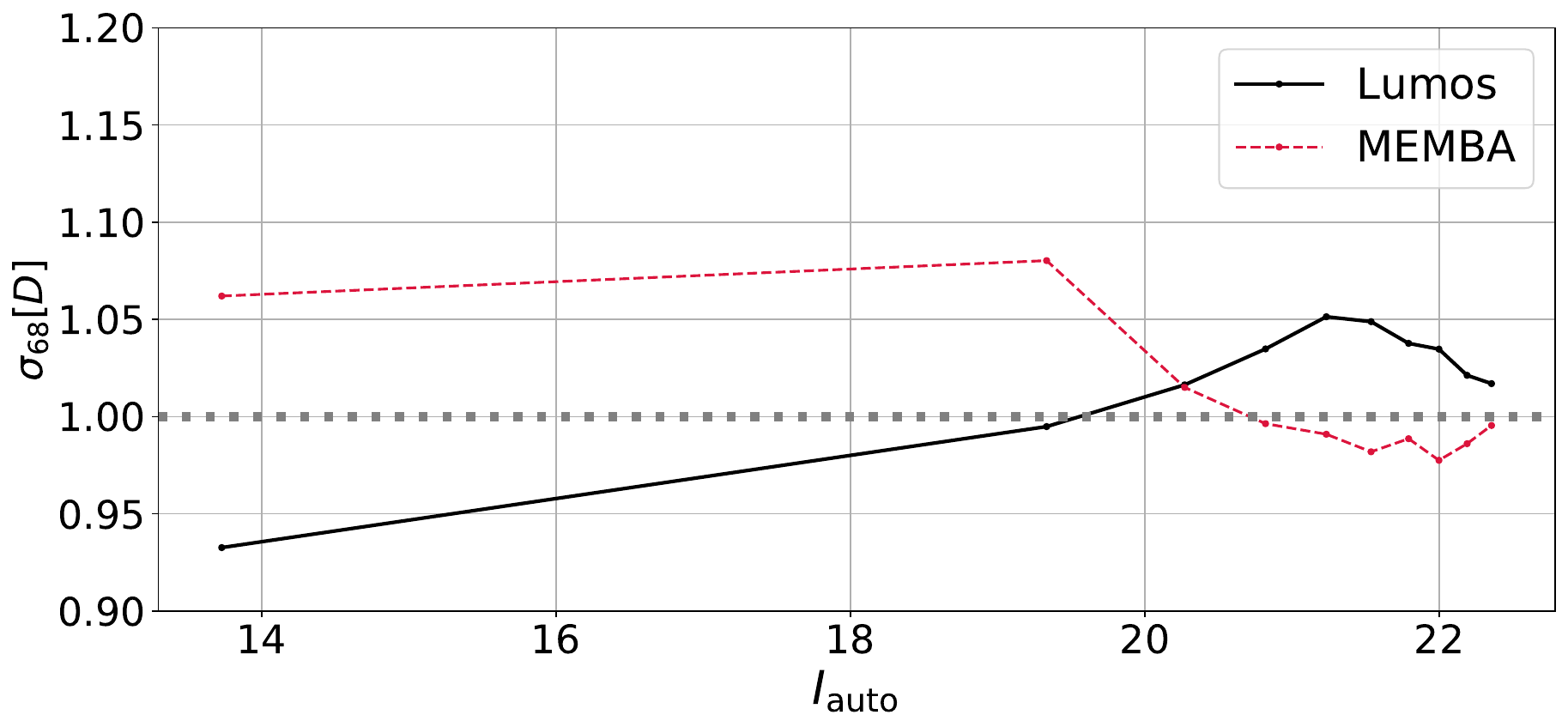}
\centering
\caption{The width of the distribution in Equation \ref{eq:Duplicates} for \textsc{Lumos} and \textsc{MEMBA} flux predictions in equally populated magnitude bins. Robust uncertainties must provide a unity width (marked by the thick grey dotted line).}
\label{fig:duplicates}
\end{figure}

Figure \ref{fig:duplicates} shows the width of the $D$ distribution in equally populated magnitude bins.  If the photometry uncertainties are properly accounted for, the distribution of $D$ (Eq. \ref{eq:Duplicates}) should be a Gaussian with unit standard deviation. To be less affected by outliers, we have estimated the width of D with $\sigma_{\rm 68}$ (Eq. \ref{eq:sigma68}) and
both \textsc{Lumos} and \textsc{MEMBA} display a quite constant unity $\sigma_{\rm 68}[D]$ along the tested magnitude range (solid black and red dashed lines, respectively). In the case of \textsc{MEMBA}, the background estimation with aperture photometry was providing 20\% underestimated errors at the bright end. This trend was fixed using \textsc{BKGnet} (see Fig. 11 in \cite{bkgnet}). \\

\begin{figure}
\includegraphics[width=0.47\textwidth]{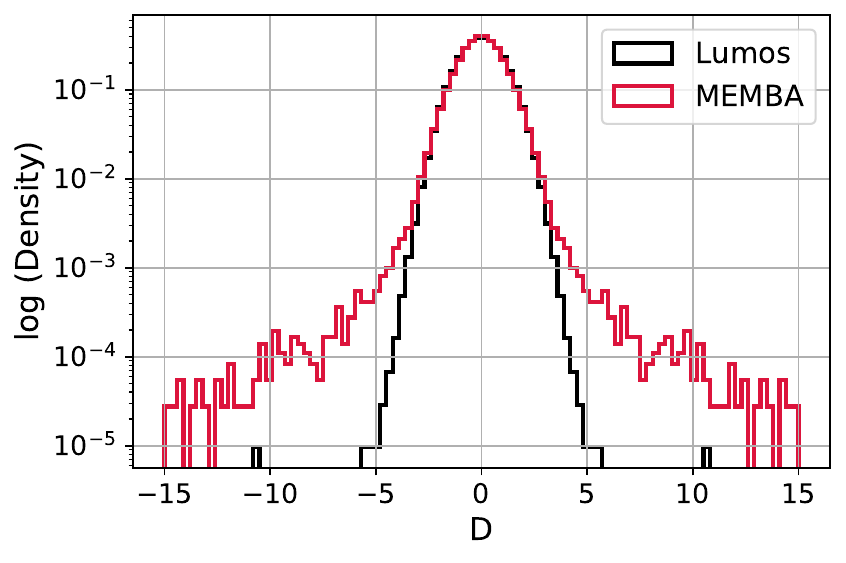}
\centering
\caption{Distribution of D (Eq. \ref{eq:Duplicates}) estimated with \textsc{Lumos} (solid black) and \textsc{MEMBA} (dashed red) photometries in logarithmic scale.}
\label{fig:duplicates_hist}
\end{figure}

Figure \ref{fig:duplicates_hist} shows the distribution of the quantity defined in Equation \ref{eq:Duplicates} with \textsc{Lumos} (solid black) and \textsc{MEMBA} (dashed red) photometries. As expected, both of them fit a Gaussian with zero mean and unit variance, however we can note  a tail of outliers in the \textsc{MEMBA} photometry not present with \textsc{Lumos}. This can be connected to \textsc{Lumos} providing not purely statistical uncertainties.
While inaccuracies in the profile parameters or contaminating effects at the image level are not considered in \textsc{MEMBA} flux errors, \textsc{Lumos} is flexible enough to provide an error estimate that already takes into account these effects. As an example, if the parent catalogue provides a 10\% underestimated $r_{\rm 50}$ for a particular galaxy, with aperture photometry the aperture size and consequently the flux measurement will be also underestimated. However, the flux uncertainty will only account for the statistical variation in the pixels within the aperture, while the error in the galaxy profile will not be considered.
In contrast, \textsc{Lumos} is provided with the galaxy and the galaxy modelled image and therefore, differences between these two are captured and accounted for in the flux uncertainty.\\

\begin{figure}
\includegraphics[width=0.47\textwidth]{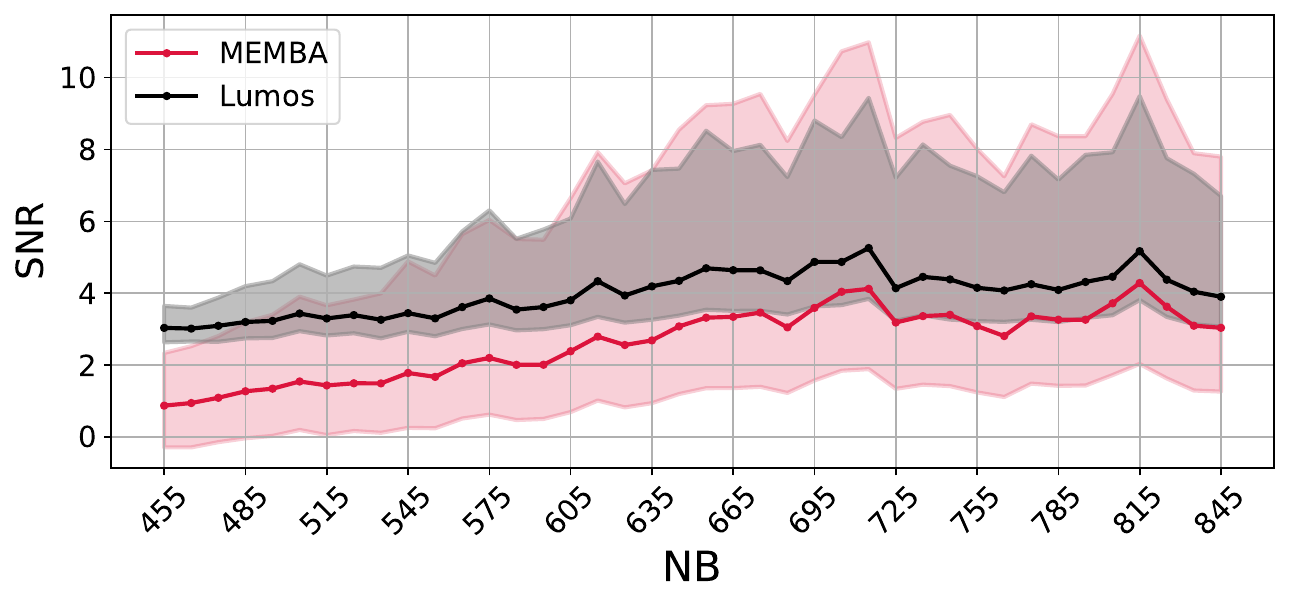}
\centering
\caption{Median SNR per narrow band filter with \textsc{Lumos} and \textsc{MEMBA} flux measurements. Shaded areas are the 16-th and 84-th quantiles of the SNR distribution.}
\label{fig:SNR}
\end{figure}

Figure \ref{fig:SNR} compares the median SNR per narrow band in \textsc{MEMBA} and \textsc{Lumos} photometries for galaxies with $i_{\rm AB} < 22.5$. The shaded areas correspond to the 16-th and 84-th quantiles of the SNR distribution. For the complete photometry catalogue, on average \textsc{Lumos} provides a 54\% higher SNR. Furthermore, it gives a higher median SNR at all wavelengths, although the increment with respect to \textsc{MEMBA} is higher in bluer bands.  For galaxies with  $i_{\rm AB}>22$, the SNR is 2.5 times higher in \textsc{Lumos}. The ratio increases to 3 taking into account only the bluest narrow band ("NB455") and decreases to a factor of 2 for the reddest one ("NB845"). This is natural considering that \textsc{Lumos} gives the greatest improvement in terms of SNR for faint objects.  Altogether, $\approx 85\%$ of the observations have  higher SNR with \textsc{Lumos} photometry.\\

\subsubsection{Observation's flagging}

\begin{figure}
\includegraphics[width=0.47\textwidth]{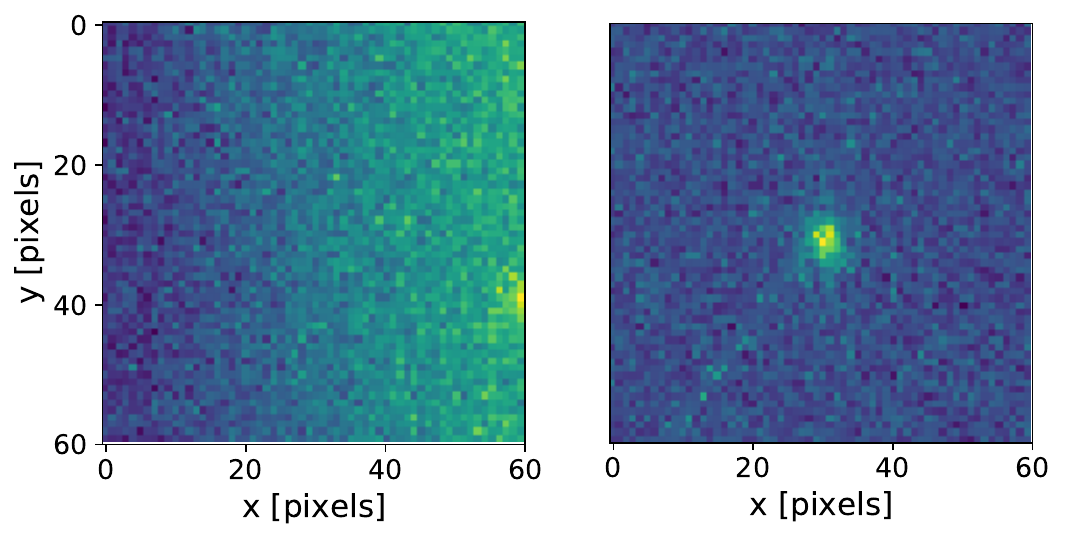}
\centering
\caption{Observations affected by scattered light (left panel) and cosmic rays (right panel). While \textsc{MEMBA} provides outlier flux measurements for both observations, \textsc{Lumos} estimates a flux close to that measured in other exposures of the same object.}
\label{fig:example_outliers}
\end{figure}

The \textsc{MEMBA} pipeline already provides an outlier flag for its measurements. This is a discrete  value that flags objects with problematic image reductions, e.g. saturated pixels, crosstalk, cosmetics, distortion or undesirable artifacts near the target source such as scattered light, cosmic rays and blending. \textsc{Lumos} uses the reduced PAUCam images, which are affected by all these effects. However, as we already showed in \citet{bkgnet} and earlier in this paper (\S\ref{sec:deblending}), \textsc{Lumos} deals with recurrent problems as scattered light or blending. Figure \ref{fig:example_outliers} shows examples of scattered light (left panel) and  cosmic ray (right panel) affected observations. For the former, \textsc{MEMBA} provides a flux of -43.87 $e^{\rm -}$/s, while \textsc{Lumos} measures 23.01 $e^{\rm -}$/s. In the cosmic ray example, \textsc{MEMBA} provides a calibrated flux of -211.21 $e^{\rm -}$/s, while with \textsc{Lumos} this is 110.76 $e^{\rm -}$/s. Other observations of the same galaxy in the same NB filter provide a mean flux 100.75 $e^{\rm -}$/s, which suggests that the \textsc{Lumos} measurement is closer to the correct flux.\\

Currently with aperture photometry 10\% of the observations are flagged. Within these flagged objects, 72\% are observations affected by scattered light, 18\% have image distortion effects and the rest is distributed among other minority effects such as e.g. crosstalk, cosmic rays or cosmetics. We have observed that \textsc{Lumos} predictions are only affected in the presence of image distortions, cosmetics and saturated pixels, which reduces the number of flagged observations from 10\% to  2\%. This reduction highlights that \textsc{Lumos} is more robust towards outliers in the photometry, which it is particularly interesting since the network is not explicitly trained to deal with artifacts as cosmic rays or crosstalk signals. However, by using real PAUCam background cutouts, we include examples of such effects in the training sample from which \textsc{Lumos} learns to make robust predictions. As a result, \textsc{Lumos} increases the size of the galaxy sample that is considered reliable. 

\subsection{Comparison with SDSS spectroscopy}
\label{sec:pausresults:sdss}
To further validate the flux estimates, we have compared our measurements with synthetic PAUS photometry. These are constructed convolving SDSS galaxy spectra with the PAUCam filter throughput. Unfortunately, the synthetic PAUS data corresponds to a bright sample with a magnitude limit $i_{\rm AB} < 20.5$, which only provides validation of  bright sources. Comparing PAUS with PAUS synthetic data requires having spectra and PAUS photometry of the same galaxies and matching them by sky position (we have paired galaxies within 0.5 arcsec). It also requires scaling the synthetic  PAUS fluxes with a multiplicative zero point (\rm{\rm{zp}}). The zero point is obtained by minimising the $\chi^2$ between PAUS observations and PAUS synthetic fluxes, i.e.
\begin{equation}
    \chi^2 = \sum_{\rm i} \frac{\left(f_{\rm SDSS,i} -  {\rm zp} \cdot f_{\rm SDSS_{\rm PAUS, i}}\right)^2}{\sigma_{\rm SDSS,i}^2 + {\rm zp}^2 \cdot
    \sigma_{\rm SDSS_{\rm PAUS,i}}^2},
    \label{eq:zp_SDSS} 
\end{equation} where SDSS is the observed SDSS photometry,
${\rm SDSS}_{\rm PAUS}$ is the SDSS-PAUS synthetic flux and the sum (i) is over the $gri$ bands. This minimisation provides a median  of 1.64 with a $\sigma_{\rm 68}$ = 1.01. \\

\begin{figure}
\includegraphics[width=0.47\textwidth]{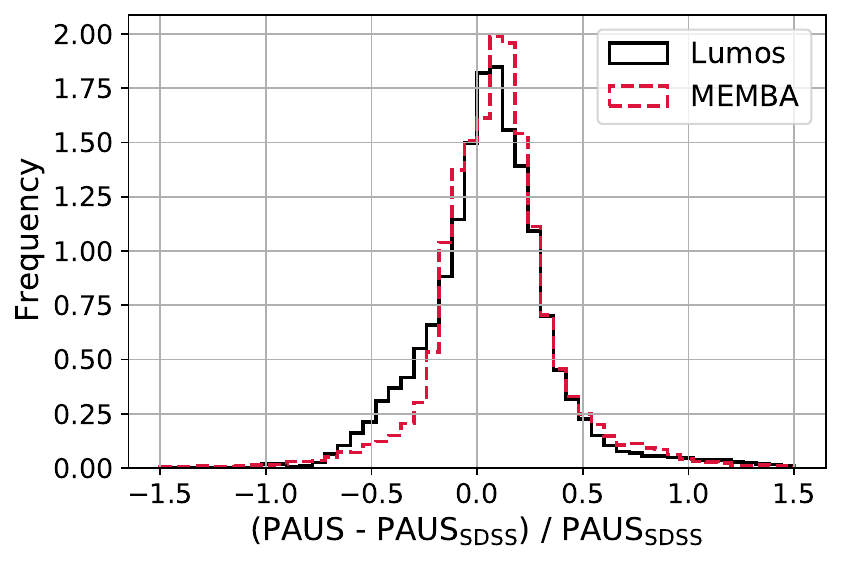}
\centering
\caption{Comparison between PAUS flux measurements and SDSS measurements convolved with PAUCam filters (PAUS synthetic fluxes). The solid black line corresponds to \textsc{Lumos} measurements and the red dashed line to \textsc{MEMBA}.}
\label{fig:SDSS_comp}
\end{figure}

Figure \ref{fig:SDSS_comp} shows that both \textsc{MEMBA} and \textsc{Lumos} agree well with SDSS-PAUS convolved flux measurements. \textsc{MEMBA} displays a lower spread than \textsc{Lumos}, with $\sigma_{\rm 68}$ = 0.22 and 0.24, respectively. Nonetheless, \textsc{Lumos} shows a 5\% bias while in \textsc{MEMBA} this goes to 10\%. Furthermore, the number of observations at more than $5\sigma$ from the mean of the distribution is reduced by 2 with \textsc{Lumos}, going from 3\% to 1.5\%.

\subsection{Coadded flux measurements}
\label{sec:pausresults:coadds}
The co-added flux measurements are constructed combining individual observations of the same galaxy in the same narrow band.  Co-adding exposures increases the SNR of the galaxy photometry and it is very helpful to reject wrong observations. Before co-adding individual observations, a  zero-point calibration per image is required, which in our case is done relative to SDSS  (see \S\ref{sec:pausdata} for more details about PAUS data). One common way of co-adding flux measurements ($f_{\rm coadd}$), and the currently implemented in PAUS, is a weighted sum of the individual observations

\begin{equation}
    f_{\rm coadd} = \frac{\sum_{\rm i} f_{\rm i} / \sigma_{i}^2} {\sum_{\rm i} 1/\sigma_{i}^2},
    \label{eq:coadds_indexp}
\end{equation} 
where the weights are the inverse variance of the observations and $f_{\rm i}$ and $\sigma_{i}^2$ are the flux measurement and its variance of the i-th observation, respectively. \\

\begin{figure}
\includegraphics[width=0.47\textwidth]{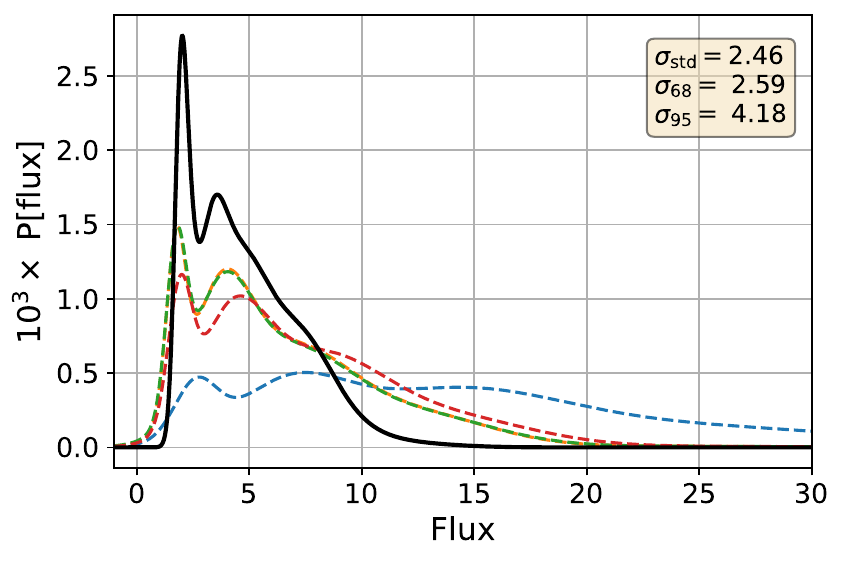}
\centering
\caption{The co-added flux probability distribution (black solid line) constructed from its individual observations (colored dashed lines). The upper box displays the $\sigma_{\rm std}$, $\sigma_{\rm 68}$ and $\sigma_{\rm 95}$ of the co-added PDF.}
\label{fig:coadd_pdf}
\end{figure}

\textsc{Lumos} calibrates the Gaussian components individually with the photometric zero point in such a way that when the these are combined, they already provide a calibrated PDF for the flux observation.  With \textsc{Lumos}, combining point like estimates with Equation \ref{eq:coadds_indexp} is still possible. Nevertheless, it can also generate co-added measurements combining  the probability distribution of the individual galaxy observations. Figure \ref{fig:coadd_pdf} shows an example of the co-added flux PDF of a faint galaxy. The dashed coloured lines correspond to the individual observations while the black line is the co-added PDF. This example also shows the benefit of creating co-adds at a PDF level. Combining point-like values would only provide a flux measurement close to the co-added PDF peak. In contrast, combining the PDFs keeps the contributions from the secondary peaks and the tails and therefore,  it contains more valuable information about the measurement than a single point-like estimate.\\

Furthermore, having the PDF allows the calculation of the flux and flux error using different statistical estimators, e.g. the median or the peak. From the co-added flux PDF, we estimate the mean, median, the peak, the variance, $\sigma_{\rm 68}$ and $\sigma_{\rm 95}$.  From the two latest quantities, we construct a measurement of the PDFs gaussianity,
\begin{equation}
    \eta \equiv  \sigma_{\rm 95} / \sigma_{\rm 68} -1\, ,
    \label{eq:gaussianity}
\end{equation} which can be helpful to decide which are the best flux and flux error estimators. While for a sharp and peaked  PDF ($\eta \gtrapprox 1$), the peak or the median would provide similar flux estimates, in non-gaussian PDFs as e.g. the galaxy in Figure \ref{fig:coadd_pdf} these two estimators would give significantly different measurements. The PDF gaussianity also affects the flux uncertainty estimators. Broadly, galaxies with $\eta > 1$ have $\sigma_{\rm 68}<\sigma_{\rm std}$, while this is the opposite for galaxies with $\eta < 1$. Figure \ref{fig:coadd_pdf} ($\eta = 0.56$) is also an example of multiple peaked PDF where $\sigma_{\rm 68}$ > $\sigma_{\rm std}$ (see the top right box in the Figure for more details).\\

We have tested applying different flux and flux error estimators based on the  $\eta$ parameter. However, at the end of the day we have found that the peak of the flux PDF is the best estimator regardless of the PDF gaussianity and that using $\sigma_{\rm 68}$, $\sigma_{\rm std}$ or $\sigma_{\rm 95}/2$
does not lead to a significant difference.

\subsection{Photometric redshift estimates}
\label{sec:photoz}
Accurate photo-z estimates are crucial for many science applications. Improving the photometry SNR is expected to improve the photo-z estimates. In this section, we have tested \textsc{Lumos} photometry with  \textsc{BCNz2} \citep[][]{photoz-Martin} and  \textsc{Deepz} \citep[][]{Deepz}: a template based method and a deep learning algorithm  built specifically for estimating PAUS photo-zs. As there is not a sample of galaxies with known photometry, testing the photo-zs has also been particularly helpful to find and fix some issues in the \textsc{Lumos} photometry that were missed with other validation tests. One example of this are galaxies which were exhibiting an oscillating photometry. This kind of objects were detected as photo-z outliers and we could trace that these were triggered by galaxy images with sub-pixels shifts with respect to the center of the stamp. \\

\begin{figure*}
\centering
\includegraphics[width=0.49\textwidth]{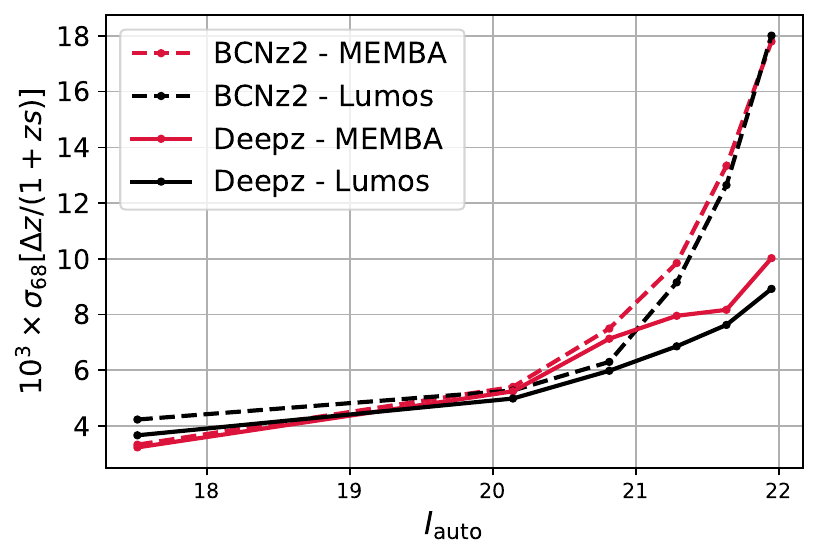}
\includegraphics[width=0.49\textwidth]{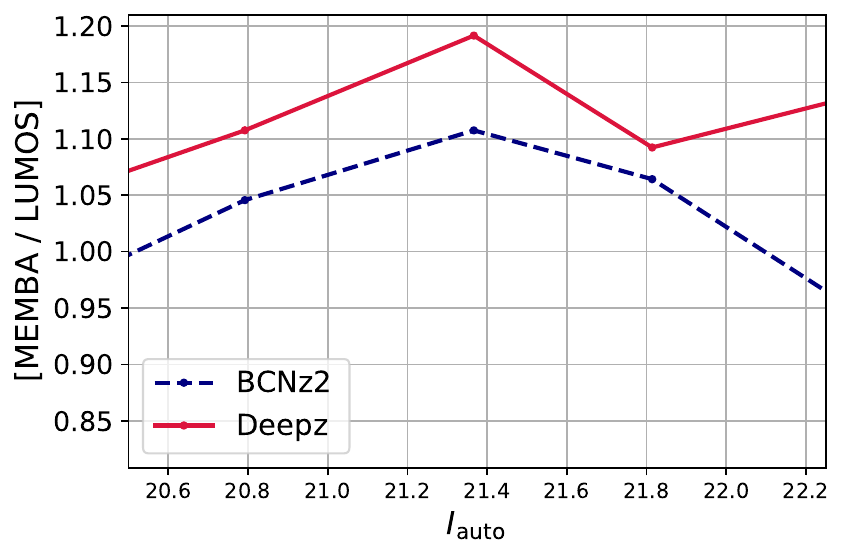}
\caption{\emph{Left:} Photo-z precision with \textsc{BCNz2} and \textsc{Deepz} using \textsc{Lumos} or \textsc{MEMBA} photometry. \emph{Right}: Relative difference between photo-zs with \textsc{Lumos} or \textsc{MEMBA} photometry.}
 \label{fig:photozs}
\end{figure*}

Figure \ref{fig:photozs} shows the photo-z dispersion  with \textsc{BCNz2} or \textsc{Deepz}  using \textsc{Lumos} photometry to $i_{\rm AB} <22.5$. We have also included the photo-z result with the \textsc{MEMBA} forced aperture photometry as a comparison.
With \textsc{BCNz2},  \textsc{Lumos} photometry reduces the photo-z scatter by  5-15\% for galaxies with $i_{\rm AB} >20.5$. However, the right panel also shows a small degradation at the faintest galaxies with  \textsc{BCNz2} on \textsc{Lumos} photometry (dashed blue line). This degradation is related with the galaxy redshift rather than to its brightness. At high redshift, photo-zs with \textsc{Lumos} photometry are statistically better, however there are some high redshfit outliers that increase $\sigma_{\rm 68}$. Rejecting galaxies with spectroscopic redshift ($z_{\rm s}$) $z_{\rm s}$ > 0.8, the faintest galaxies ($i_{\rm AB} >22$) have a 14\% lower photo-z dispersion with \textsc{Lumos} than with \textsc{MEMBA} photometry. \\

Photo-zs with \textsc{Deepz} are not showing this degradation at high redshift. For galaxies with $i_{\rm AB} >20.5$, the \textsc{Deepz} photo-zs are between 10\% and 20\% more precise with \textsc{Lumos} photometry. Furthermore, at $i_{\rm AB} >22$ and without any redshift cut, photo-zs are 15\% better. This suggests that the minor degradation with \textsc{BCNz2} at high redshift is caused by the photo-z code. \\

With both photo-z codes, the performance is degrading at the brightest end with the \textsc{Lumos} photometry. This could potentially be triggered by differences between the \sims image simulations and the data. Bright galaxies with higher SNR are more resolved. Therefore, discrepancies between the training simulations and the data are more evident and these could have a stronger effect on the network's performance. Nevertheless, the fraction of objects affected by this effect is small and furthermore, these are the brightest galaxies, which are not those we are more interested in. \\

In PAUS, a galaxy is considered an outlier if
\begin{equation}
    |z_{\rm p} - z_{\rm s}|\ /\ (1 + z_{\rm s}) > 0.02\, ,
\end{equation}where $z_{\rm p}$ is the photo-z and $z_{\rm s}$ is the spec-z. This outlier definition is very strict compared to broad band photometry, where a common outlier definition is $|z_{\rm  p} - z_{\rm s}| > 0.15\,(1 + z_{\rm s})$, e.g. \citet{CFHT_photoz,KiDs_photoz}.
\textsc{Lumos} photometry reduces the outlier rate with both \textsc{BCNz2} and \textsc{Deepz}. With \textsc{BCNz2}, the outlier rate is reduced by 5\% in the complete catalogue. With \textsc{Deepz}, the improvement is greater, with 20\% less outliers. This number increases to  23\% for objects with $i_{\rm AB}>22$.\\

To the best of our knowledge, we do not know  about any photo-z code that could properly deal with flux PDFs and both \textsc{BCNz2} and \textsc{Deepz} require point estimates for the flux and its uncertainty. Here, these quantities are estimated as the peak and $\sigma_{\rm 68}$ of the co-added flux PDF. We have also tested other quantities as e.g. the median and the standard deviation or choosing  different estimators based on the $\eta$ parameters (Eq. \ref{eq:gaussianity}). However, the peak and $\sigma_{\rm 68}$ are those providing the best photo-z estimates. \\

The photo-z improvement obtained with \textsc{Lumos} photometry is lower than expected considering the increment in the SNR. In 
Appendix \ref{App:bcnzflagship}, we have used PAUS simulated mocks to test \textsc{BCNz2} performance with SNR and its behaviour with artificially injected issues in the sample photometry. The results suggest that the photo-z improvement should be $\approx$90\% greater than what we actually see on data if the sample had perfect photometry. However, errors in the zero-point calibration and outliers in the flux measurements rapidly degrade the photo-z performance, suggesting that currently these are potential limiting factors of the photo-z performance.

\section{Conclusions and discussion}
\label{sec:conclusions}
Accurate galaxy photometry is a key ingredient for imaging surveys to obtain precise photometric redshifts. We have developed \textsc{Lumos}, a deep learning method to estimate the galaxy flux for astronomical images. \textsc{Lumos} is the evolution of \textsc{BKGnet}, a deep learning method that predicts the background light of astronomical images with strongly varying noise patterns. In contrast, \textsc{Lumos} predicts the  background subtracted galaxy flux, which requires an intrinsic background noise measurement. The algorithm has been developed for PAUCam images. \\

\textsc{Lumos} is trained on \sims galaxies, image simulations specially built for this work (see Fig. \ref{fig:PAUSsims}). \sims galaxy images use PAUCam image cutouts for the background noise. First, astronomical images contain distorting effects, e.g. scattered light or crosstalk, and artifacts like e.g. cosmic rays or blended galaxies, that trigger inaccuracies in the photometry. Including real PAUCam cutouts in our simulations ensures that \textsc{Lumos} has training examples to learn how to deal with such effects. Without explicitly developing \textsc{Lumos} to provide photometry in the presence of distorting artifacts, the network provides reliable flux measurements on PAUCam observations affected by scattered light, cosmic rays or other contaminating effects that require flagging with aperture photometry. \\

Furthermore, we have tested the \textsc{Lumos} deblending capability on simulations (see \S\ref{sec:deblending}). Without explicitly including blended galaxies in the training sample, \textsc{Lumos} is able to extract the target galaxy photometry much better than aperture photometry  (Fig \ref{fig:deblending}). While aperture photometry provides a catastrophic flux measurement for blended sources, \textsc{Lumos} is able to provide a flux with 2-10\% accuracy, depending on the distance in pixels to the overlapping source. This is particularly interesting since \textsc{Lumos} has not been written to deblend galaxies, however this came without additional cost by using deep learning and real PAUCam background noise patterns. \\

\textsc{Lumos} consists on a CNN followed by a MDN (Fig. \ref{fig:lumos}). While most photometry algorithms provide a flux value and an associated uncertainty, \textsc{Lumos} outputs the flux probability distribution as a linear combination of five  Gaussian distributions. Even if many science applications require photometry point estimates, having the PDF enables the generation of a co-added flux PDF (\S\ref{sec:pausresults:coadds}). The co-added flux PDF keeps valuable information about the individual flux exposure distributions that would be missed by combining point-like estimates. While using the full PDF would require reworking of the pipelines using the photometry as an input, this can also be part of an end-to-end photometry machine learning pipeline that goes from images to photo-z estimates. The network could benefit from all the information available in the full PDF to provide  more precise photo-z estimates.\\ 

On PAUS observations, \textsc{Lumos} provides fluxes that differ less than 1\% from the baseline aperture photometry measurements (see Fig \ref{fig:ratios}). Concerning uncertainties, our photometry errors are 40\% lower than with aperture photometry. This translates into between 1.5 and 3 times higher SNR in \textsc{Lumos} than in  \textsc{MEMBA}, with the largest improvement at the faint end (Fig. \ref{fig:SNR}). We have run the \textsc{BCNz2} and \textsc{Deepz} codes with \textsc{Lumos} photometry, resulting in a reduction of the  photo-z scatter with both (Fig \ref{fig:photozs}). The photo-z improvement using \textsc{Lumos} photometry is greater with \textsc{Deepz} rather than with \textsc{BCNz2}, with an overall scatter reduction of 10\% on the full catalogue and 13\% for galaxies with $i_{\rm AB} >22$ and an outlier rate reduction of $\approx$ 20\%. Nevertheless, Appendix \ref{App:bcnzflagship} shows that the photo-z improvement is limited by the photometric calibration and outliers in the sample. These outliers can have different natures as e.g. the \textsc{Lumos} photometry itself or problems in the reduced PAUCam images. \\

\textsc{Lumos} obtains the largest improvement at fainter galaxies, showing less degradation than aperture photometry. Future imaging surveys like {\it Euclid} or Rubin will observe much deeper galaxies with very low SNR, where \textsc{Lumos} could be a very helpful tool to improve the photometry.  Furthermore, \textsc{Lumos} has been shown to be robust for blended sources, which could be  also beneficial for future deeper surveys where the number of blended galaxies will significantly increase.\\

Although we have only tested the method on PAUCam images so far, we believe the methodology should readily apply to other imaging surveys.
Training the network for other surveys can be addressed by training the network from scratch with simulated galaxies using real background cutouts from the targeted survey. Nevertheless, one potential difficulty of the method applied to deeper surveys is the the modelled galaxy profile input requirement. While PAUS galaxies are shallow and therefore, there exists previous observations, deeper surveys like Rubin will observe galaxies for which there is not previous knowledge. Not using the modelled profile in the training barely affects the overall predicted flux measurements, however this degrades the SNR by 15\%.\\

\textsc{Lumos} supersedes \textsc{BKGnet} and provides a background subtracted flux measurements, which requires a measure of the background light contribution. Consequently, \textsc{Lumos} deals with potential correlations between the galaxy flux and the background light that are not easy to address analytically. Moreover, in this work we have combined two independent networks, \textsc{Lumos} and \textsc{Deepz}, which provides the greatest photo-z obtained. This motivates the construction of an end-to-end pipeline that supersedes \textsc{Lumos} providing the galaxy photometry and the photometric redshift.

\section*{Acknowledgements}
The authors thank Malgorzata Siudek, Pablo Renard, Giorgio Manzoni, Jacobo Asorey, Luca Tortorelli and Helena S\'anchez-Dom\'inguez for providing feedback on the paper.\\
The PAU Survey is partially supported by MINECO under grants CSD2007-00060, AYA2015-71825, ESP2017-89838, PGC2018-094773, PGC2018-102021, SEV-2016-0588, SEV-2016-0597, MDM-2015-0509, PID2019-111317GB-C31 and Juan de la Cierva fellowship and LACEGAL and EWC Marie Sklodowska-Curie grant No 734374 and no.776247 with ERDF funds from the EU Horizon 2020 Programme, some of which include ERDF funds from the European Union. IEEC and IFAE are partially funded by the CERCA and Beatriu de Pinos program of the Generalitat de Catalunya. Funding for PAUS has also been provided by Durham University (via the ERC StG DEGAS-259586), ETH Zurich, Leiden University (via ERC StG ADULT-279396 and Netherlands Organisation for Scientific Research (NWO) Vici grant 639.043.512), Bochum University (via a Heisenberg grant of the Deutsche Forschungsgemeinschaft (Hi 1495/5-1) as well as an ERC Consolidator Grant (No. 770935)), University College London, Portsmouth support through the Royal Society Wolfson fellowship and from the European Union's Horizon 2020 research and innovation programme under the grant agreement No 776247 EWC.  \\
The PAU data center is hosted by the Port d'Informaci\'o Cient\'ifica (PIC), maintained through a collaboration of CIEMAT and IFAE, with additional support from Universitat Aut\`onoma de Barcelona and ERDF. We acknowledge the PIC services department team for their support and fruitful discussions. We gratefully acknowledge the support of NVIDIA Corporation with the donation of the Titan V  GPU used for this research.

\section*{Data availability}
The PAUS raw data is publically available through the ING group. A few reduced images are publically available at https://www.pausurvey.org.

\bsp	
\label{lastpage}

\bibliographystyle{mnras}
\bibliography{flux}

\begin{appendices}
\section{Flux estimation methods: derivations}
\label{App:flux_methods}
This appendix derives the linear combination of pixel values giving and unbiased and optimal flux measurement (Eq. \ref{eq:optweights}).
The SNR of the measurement when combining pixels with mean $m$ and weight $w$ is

\begin{equation}
    {\rm SNR} = \frac{\sum_{\rm i} w_{\rm i} m_{\rm i}}{\sqrt{\sum_{\rm i} w_{\rm i}^2 (m_{\rm i} + b_{\rm i})}},
    \label{SNR}
\end{equation} where $b_i$ is the background mean value. The optimal SNR is found by requiring stationary derivatives for all weights independently, which results in

\begin{equation}
w_{\rm x} = \lambda \frac{m_{\rm x}}{(m_{\rm x}+b_{\rm x})} 
\label{eq_weight}
\end{equation}

\noindent
where $\lambda$ is a constant. The flux
measurement being unbiased means
\begin{equation}
    \sum_{ \rm i} m_{\rm i} = \sum_{\rm i} w_{\rm i} m_{\rm i}.
\label{eq_unbiased}
\end{equation} 

\noindent
Using this requirement, the pixel weights (Eq.\ref{eq_weight}) becomes
\begin{equation}
    w_{\rm x} =  \frac{\sum_{\rm i} m_{\rm i}}{\sum_{\rm i} m_{\rm i}^2/(m_{\rm i}+b_{\rm i})} \frac{1}{1+b_{\rm x}/m_{\rm x}}
\end{equation}
Notice that, given a pixel x, its weight $w_{\rm x}$ depends on the true flux ($m_{\rm x}$) and background ($b_{\rm x}$) on that concrete pixel. 

\section{Convolutional Neural Networks}
\label{App:CNNs}
Machine learning methods are data analysis techniques where the algorithm learns from the data. In particular, one of the currently most popular classes of algorithms are neural networks \citep{lecun98}, which are designed to recognise patterns, usually learned from training data (\textit{supervised method}). They are mainly used for regression and classification problems. Deep learning is a subset of machine learning that refers to the development of neural network technology involving a large number of layers. \\

Other terms that one needs to be familiar with are \textit{epoch} and \textit{batch}. An \textit{epoch} is an iteration over the complete training dataset. However, it is common practice to give the data to the network in \textit{batches}. Feeding the network in batches helps it learn faster as in every iteration over a batch, it updates all the weights. Then, instead of updating once per epoch, it updates as many times as there are batches, which speeds up the process. \\

Deep learning methods, and in general any supervised machine learning algorithm, model a problem by optimising a set of trainable parameters that fit the data. This is done in three stages: \textit{forward propagation}, \textit{back propagation} and \textit{weight optimisation}. The network starts with the \textit{forward propagation}. At this stage, the input data propagates through all the network layers to give a prediction for each of the input samples. After that, by comparing the prediction with the known true value (\textit{label}), the network estimates a prediction error with a loss function.  The ultimate goal of the training procedure is to minimise the loss function. \\

To minimise the loss, one performs a process named \textit{back propagation} after evaluating the loss function. \textit{Back propagation} consists of computing the contribution of each weight to the loss function \citep{backpropagation}. Such contributions are calculated using the chain rule. The \textit{weight optimisation} consists in updating the network parameters using the derivatives estimated. This whole procedure takes place repeatedly, reducing the loss function after each iteration while adapting the parameters to the data.
The amount of variation allowed per iteration is regulated by the \textit{learning rate}.\\

After each  layer there is an activation function, which are non-linear functions that map the outcome of a layer to the input of the following one. This is required to produce non-linearities in the model. An example of an activation function is the Rectified Linear Unit (ReLu) \citep{Alexnet}.\\

In this work, we use a Convolutional Neural Network \citep[CNN;][]{lecun98}. Our network contains four differentiated types of layers:  
\begin{description}
    \item [\emph{Convolutional layer}:]  This layer makes the network powerful in image and pattern recognition tasks. It has a filter, technically named \textit{kernel} and is usually 2-dimensional, which contains a set of trainable weights used to convolve the image. The outcome of this layer is the input image convolved with the kernel. In a given convolutional layer, each of the outputs is a linear combination of the different convolutions. Each of these convolutions will generate a convolved image, which we refer to as \textit{channel}. All of them together are the input of the next layer.
    
    \item [\emph{Pooling layers}:] This layer reduces the  dimensionality of the set of convolved images. It applies some function (e.g. maximum, sum, mean) to a group of  spatially connected pixels and reduces the dimensions of such group. For example, the Max-Pooling, which we use in this paper, takes a e.g. (2x2) group of pixels and converts them to their maximum. Although we use it to handle the amount of data generated after the convolutions, it also regularises the model to avoid learning from non-generalisable noise and details in the training data (also known as overfitting). 
    \item [\emph{Batch normalisation layer}:] In this layer the network normalises the output of a previous activation layer. It subtracts the  mean and divides by the standard deviation. Batch normalisation helps to increase the stability of a neural network and avoids over-fitting problems \citep{batchnorm}.
    \item [\emph{Fully connected layer}:] These layers are usually the last layers of the network and its input is the linearised outcome of the previous ones (in our network: convolutions, poolings and batch normalisations). It applies a linear transformation from the input to the output. 

\end{description}

\section{Forecasting the effect of errors on profile parameters}
\label{App:FF}
The algorithms described in \S\ref{sec:methods} need information about the galaxy profile properties, making the flux measurement accuracy sensitive to errors on these parameters. 
Here we will quantify the effect that errors in the input galaxy parameters have on the flux measurements using a Fisher forecast formalism \citep{Fisher1922} on \sims galaxies.
In particular, we will test if
neural networks are more robust
to errors in the input parameters.\\

A photometry algorithm ($\Phi$) that measures the flux ($\tilde{f}$) 
\begin{equation}
    \tilde{f} = \Phi(I,f, r_{\rm 50}, n_{\rm s}, PSF, e, b) 
    \label{eq:flux_dependencies}
\end{equation} 

\noindent
is foremost dependent on the galaxy image ($I$), but also
parameters such as the total flux ($f$),
the half-light radius ($r_{\rm 50}$), the S\'{e}rsic index ($n_{\rm s}$), the PSF FWHM, the ellipticity ($e$) and the background light ($b$). This
is because e.g. the aperture algorithm uses
these quantities to scale the pixels and they
are used to construct a profile for the profile fit. We can estimate the error on
the flux from these parameters using a Fisher
matrix formalism.
The Fisher matrix is estimated by
\begin{equation}
    \textbf{\textsf{FM}}_{\rm \mu \nu} = \frac{\partial \tilde{f}}{\partial \mu} \left(\sigma_{\tilde{f}}^{-2}\right)\frac{\partial \tilde{f} }{\partial \nu}\, , \label{eq:fisher} 
\end{equation} 

\noindent
where the indices $\mu$ and $\nu$ are the different parameters the total flux depend on (see Eq. \ref{eq:flux_dependencies}). \\

\begin{figure}
\includegraphics[width=0.47\textwidth]{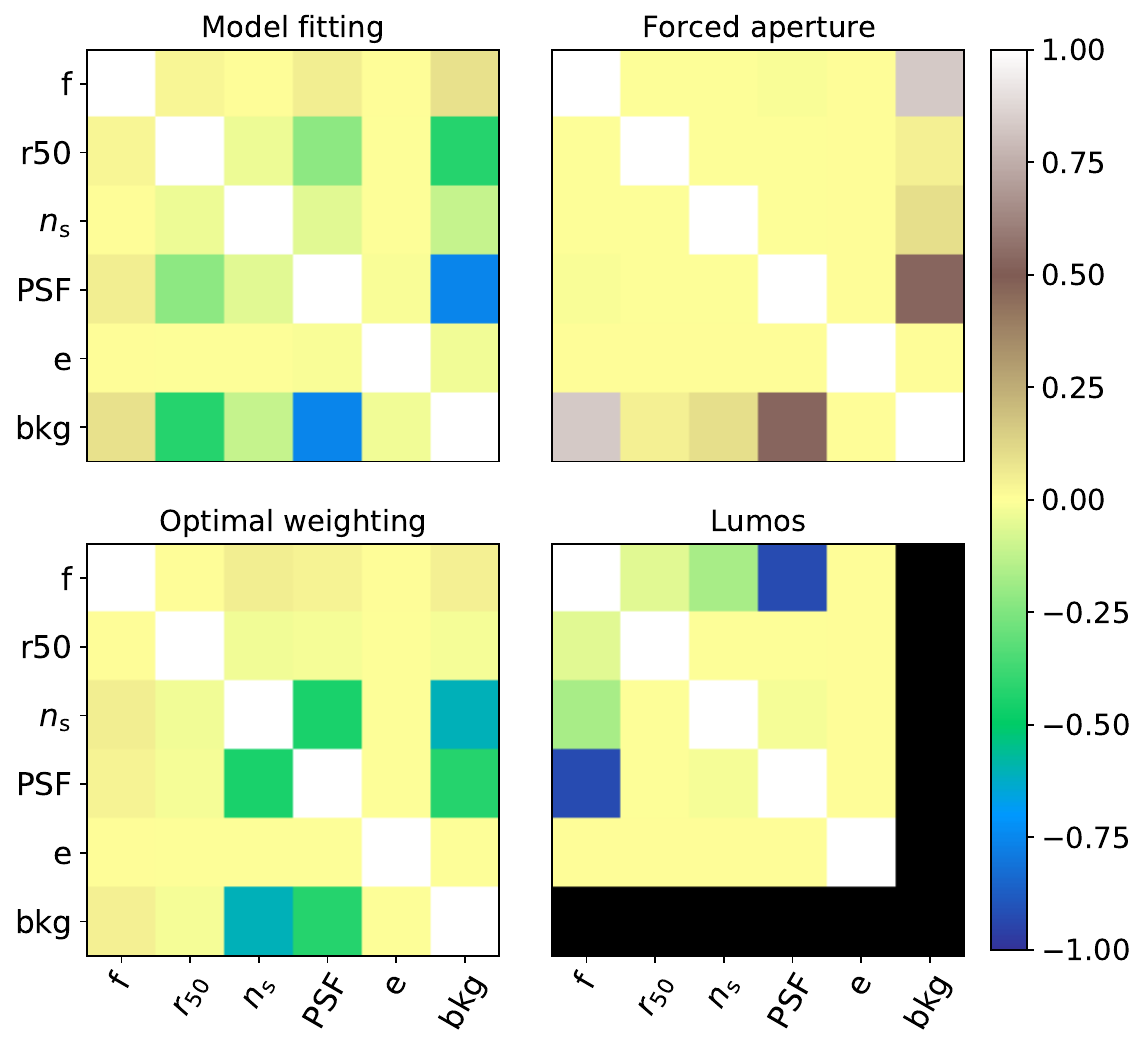}
\centering
\caption{The correlation matrix for the parameters $f,r_{\rm 50},n_{\rm s},PSF,e$ and $bkg$ for: \emph{Top left:} The model fitting method, \emph{Top right:} Forced aperture photometry, \emph{Bottom left:} Optimal weighted pixel sum and \emph{Bottom right:} \textsc{Lumos}.}
\label{fig:correlations}
\end{figure}

The covariance matrix of the flux measurements can be estimated as the inverse of the Fisher matrix.
Figure \ref{fig:correlations} shows the correlation matrices of the parameters $Flux, r_{\rm 50}, n_{\rm s}, PSF, e$ and $b$ for the four flux estimation methods described in \S\ref{sec:methods}. The correlation matrix differs for different galaxy types (e.g. different morphologies or brightness). Here we have constructed a galaxy with values of $r_{50}$, $n_{\rm s}$ and $PSF$ corresponding to the mean of their distribution.
The model-fitting method (top left panel), together with the optimal weighting (bottom left panel) are those showing more correlation between the flux and the profile parameters. In contrast, the forced aperture photometry and \textsc{Lumos} show a lower correlation. The lower correlation makes these methods more robust since the effect of uncertainties in the parameters is also lower. \\

All methods but \textsc{Lumos} show a high correlation with the background estimation.
The background light is not an input parameter for \textsc{Lumos}, since it is intrinsically measured inside the method. 
This makes \textsc{Lumos} insensitive to external errors on this parameter. \\

As mentioned, the correlation matrices in Figure \ref{fig:correlations} are for a particular common galaxy. To estimate the requirements on these parameters, we have studied how errors in the profile parameters propagate to errors in the flux measurement. For that, we have assumed a 10\% prior error in each of the input parameters, such that
\begin{equation}
    \textbf{\textsf{FM}}_{\rm comb} = \textbf{\textsf{FM}} + \textbf{\textsf{FM}}_{\rm priors},
\end{equation} where $\textbf{\textsf{FM}}_{\rm priors}$ is a diagonal matrix including the inverse prior variance of each parameter. The variance on the flux parameter is then 
\begin{equation}
\textbf{\textsf{$\sigma^2_{\rm f}$}} = {\left(\textbf{\textsf{FM}}^{\rm -1}_{\rm comb}\right)}_{\rm ff}
\end{equation}

\noindent
where the matrix subscripts (ff) denotes selecting
the row/column corresponding to the flux
parameter. \\

Table \ref{tab:fc} shows the percentage of error in the parameters that propagates to a 10\% error in the flux measurements. While studying a particular parameter, we always assume that the rest are fixed. As the sensitivity to the parameters can vary among galaxy types, the results in Table \ref{tab:fc} are averaged out of one hundred independent random galaxies. \textsc{Lumos} is the method that requires higher errors on the parameters to propagate to a 10\% flux error, i.e. it shows as the most robust method. As expected from Figure \ref{fig:correlations}, the PSF is the parameter it is more sensitive to, followed by the half light radius. However, in both cases it is still less sensitive than the other methodologies. \\

 The robustness of \textsc{Lumos} most likely comes from having the galaxy image.  \textsc{Lumos} is provided with the galaxy and the modelled profile images, which allows the comparison and detection of problematic profiles instead of blindly relying on the input parameters. Furthermore, small errors in such parameters are more subtle and difficult to detect when these are encoded in a profile rather than directly inputted into the method.

\begin{table}
\begin{tabular}{lccccc}
\multicolumn{1}{l}{}            & $r_{\rm 50}$ & $n_{\rm s}$  & \multicolumn{1}{l}{PSF} & \multicolumn{1}{l}{e} & \multicolumn{1}{l}{b} \\ \hline
Model-fitting                   & 4            & 4  & 4                       &5                               & 1                              \\ 
Forced photometry               & 21           & 59 & 19                       & 1                               & 1                              \\ 
Opt. weighted sum               & 8            & 11  & 4                       & 13                               & 1                              \\ 
 \textsc{Lumos} & 28           & 80 & 14                       & 83
& -                              \\ \hline
\end{tabular}
\caption{Percentage of error in $r_{\rm 50}$, $n_{\rm s}$, PSF, ellipticity ($e$) and background noise ($b$) that propagates to a 10\% error in the total flux. Note that when studying one of the parameters, the rest remain fixed. Also note that high errors indicate that the method is more robust, since it requires a large error in the parameter to propagate to a 10\% flux error. }
\label{tab:fc}
\end{table}

\section{Photometric redshifts with \textsc{BCNz2} on PAUS galaxy mocks}
\label{App:bcnzflagship}
In this section, we will run \textsc{BCNz2} on the PAUS galaxy mock. The purpose of this test is to have an idea of the photo-z improvement that we should expect on data. \\

We have generated PAUS photometry with the same pipeline as the Flagship simulations (Castander et al (in prep.)) containing ~500K objects over 25 $deg^{2}$ with a redshift limit of 2.25. Initially, galaxies are generated with rest-frame luminosity using abundance matching between the halo mass function and SDSS galaxies. Next, the galaxy redshift is estimated using evolutionary population synthesis models. Then, mock galaxies are matched to the COSMOS galaxies from \citet{Ilbert2008cat} and extinction and an SED are assigned to each of them. The SED templates also take into account the emission lines. $H_{\rm \alpha}$ is computed form the ultra-violet following \citet{Halpha}.  The other line fluxes are computed following observed relations. Finally, the SED is convolved with the filter transmission curves to produce the fluxes.  \\

\begin{figure}
\includegraphics[width=0.47\textwidth]{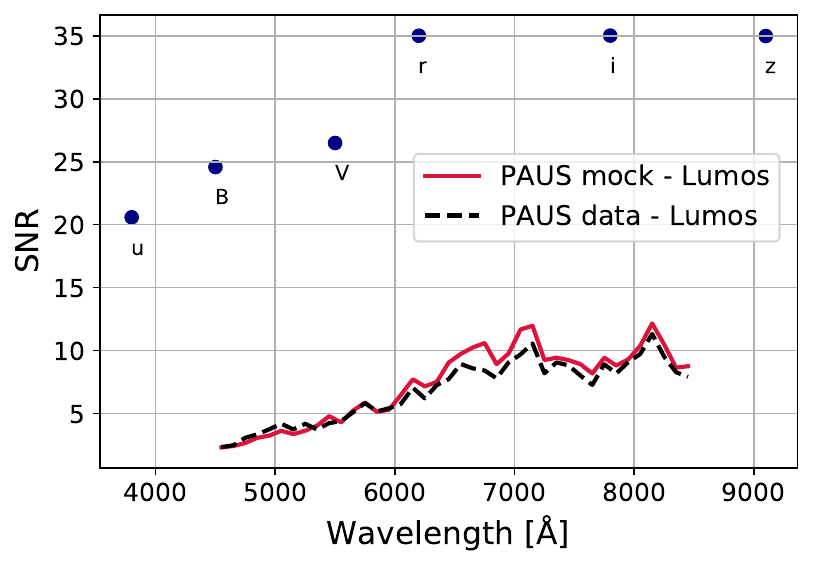}
\centering
\caption{The SNR of the PAUS galaxy simulations used to run \textsc{BCNz2}. The solid red line corresponds to a PAUS mock with a SNR similar to that provided by \textsc{Lumos}. As a comparison, the dashed black line corresponds to the observed SNR on PAUS data with \textsc{Lumos} photometry.}
\label{fig:SNR_flagship}
\end{figure}

In this work, we have used PAUS narrow bands, the CFHT $u$-band and the Subaru $BVriz$ broad bands and the noise is included assuming Gaussian uncertainties. We have randomly selected 10,000 noiseless galaxies and included uncertainties to mimic the SNR provided by \textsc{Lumos} (Fig. \ref{fig:SNR_flagship}). For the broad bands, we have generated errors to simulate the SNR in \cite{photoz-Martin}. \\

\begin{figure}
\includegraphics[width=0.47\textwidth]{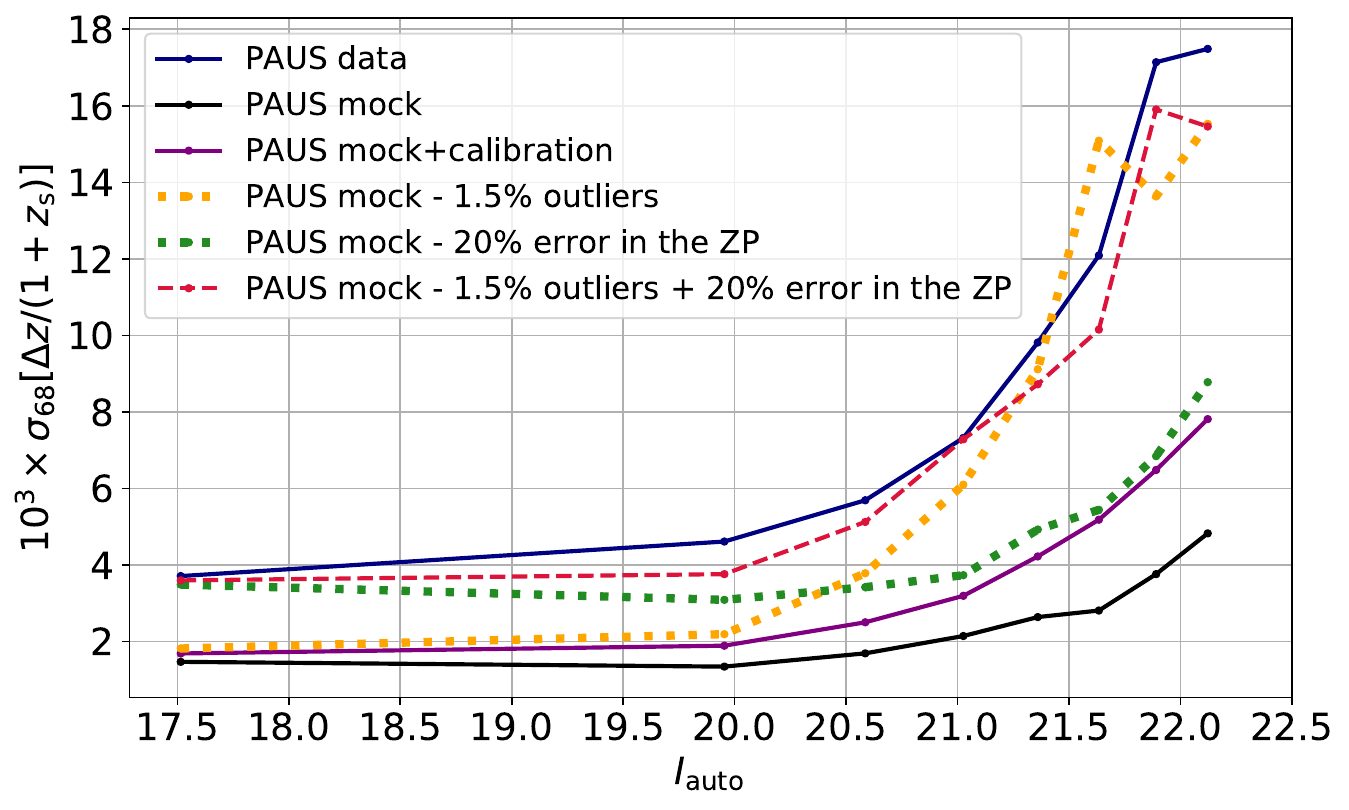}
\centering
\caption{Photo-z dispersion as a function of $i$-band magnitude using \textsc{BCNz2} for a galaxy mock with \textsc{Lumos} SNR (black solid line) and PAUS data with \textsc{Lumos} photometry (blue solid line). The purple line includes the photometric calibration on the PAUS mocks. Dotted lines include outliers and calibration errors in the PAUS mock photometry.  The red dashed line combines the effect of the two dotted lines.}
\label{fig:bcnz_flagship}
\end{figure}

Figure \ref{fig:bcnz_flagship} shows that the photo-z on the PAUS-mock (black solid line) has on average 90\% less scatter than on data (blue solid line). This is a large number considering that the SNR on both samples is similar and consequently, a similar performance is expected. Therefore, the photo-zs appear limited by other factors than the photometry SNR. \\

The purple solid line shows the effect of the photometric calibration. For each flux measurements, we have generated five individual observations resembling the number of PAUS observations in the COSMOS field from a Gaussian centered at the co-added flux.
Each of these observations is randomly assigned with a photometric zero point and its uncertainty from the distribution of zero-point in the PAUS COSMOS data. The zero points are scattered with their uncertainty, sampling for every galaxy a new zero point from a Gaussian. The individual observations are combined back to a single flux error using the scattered zero points. \\

The median zero point uncertainty in the PAUS data in COSMOS is $\approx$ 4\%. Including this effect in the photo-zs (purple solid line in Fig. \ref{fig:bcnz_flagship}) degrades the precision by 40\% with respect to perfect photometry (black solid line), from $\sigma_{\rm 68}$ = 0.0026 to $\sigma_{\rm 68}$ =  0.0041,  with faint galaxies being more affected. Nevertheless, after including the calibration effect, the photo-z scatter is still significantly lower on the PAUS mock compared to the results on data.  The coloured dotted lines in Figure \ref{fig:bcnz_flagship} shows the photo-z dispersion with further effects in the photometry that could potentially lead to a photo-z degradation. All the lines are built on the purple one, assuming the calibration effect.\\

The green dotted line studies the effect of having a addition 20\% error in 20\% of the photometric zero points. Note that this additional error is not accounted for in the final photometric error and  could potentially make a correct observation an outlier.This further reduced the photo-z precision to $\sigma_{\rm 68}$ = 0.0050, significantly affecting bright galaxies. Indeed, this effect is required to understand the photo-z degradation at the bright end on data with respect to simulations. \\

In the orange dotted line, we have artificially injected 1.5\% of outliers in the PAUS mock fluxes. These outliers are directly affecting the co-added flux measurement and therefore, a 1.5\% of affected fluxes corresponds to a higher percentage of affected galaxies.  Particularly, 45\% of the galaxies in the PAUS mock have at least one affected NB flux measurement and $\approx10\%$ have more than one. Nevertheless, mind that not all the galaxies with affected  photometries end up providing worse redshift estimates. Indeed, these outliers barely affect the bright end, where galaxies have high SNR and the photo-z algorithm deals well with an outlier in one of the bands. Contrary, at the faint end  outliers increase the photo-z scatter by $\approx$2. We have also tested the effect of other percentages of outliers finding that 1\% was too few to explain the degradation on data and 2\%, too much. With 1.5\% of outliers, the photo-z precision degrades by 80\%, providing a $\sigma_{\rm 68}$ = 0.0074. The dashed red line combines the two previous effects and gives a photo-z precision  close to that obtained on data.

\section{Colour histograms in the complete narrow  band set}
\label{App:colorhists}
\begin{figure*}
\includegraphics[width=0.97\textwidth]{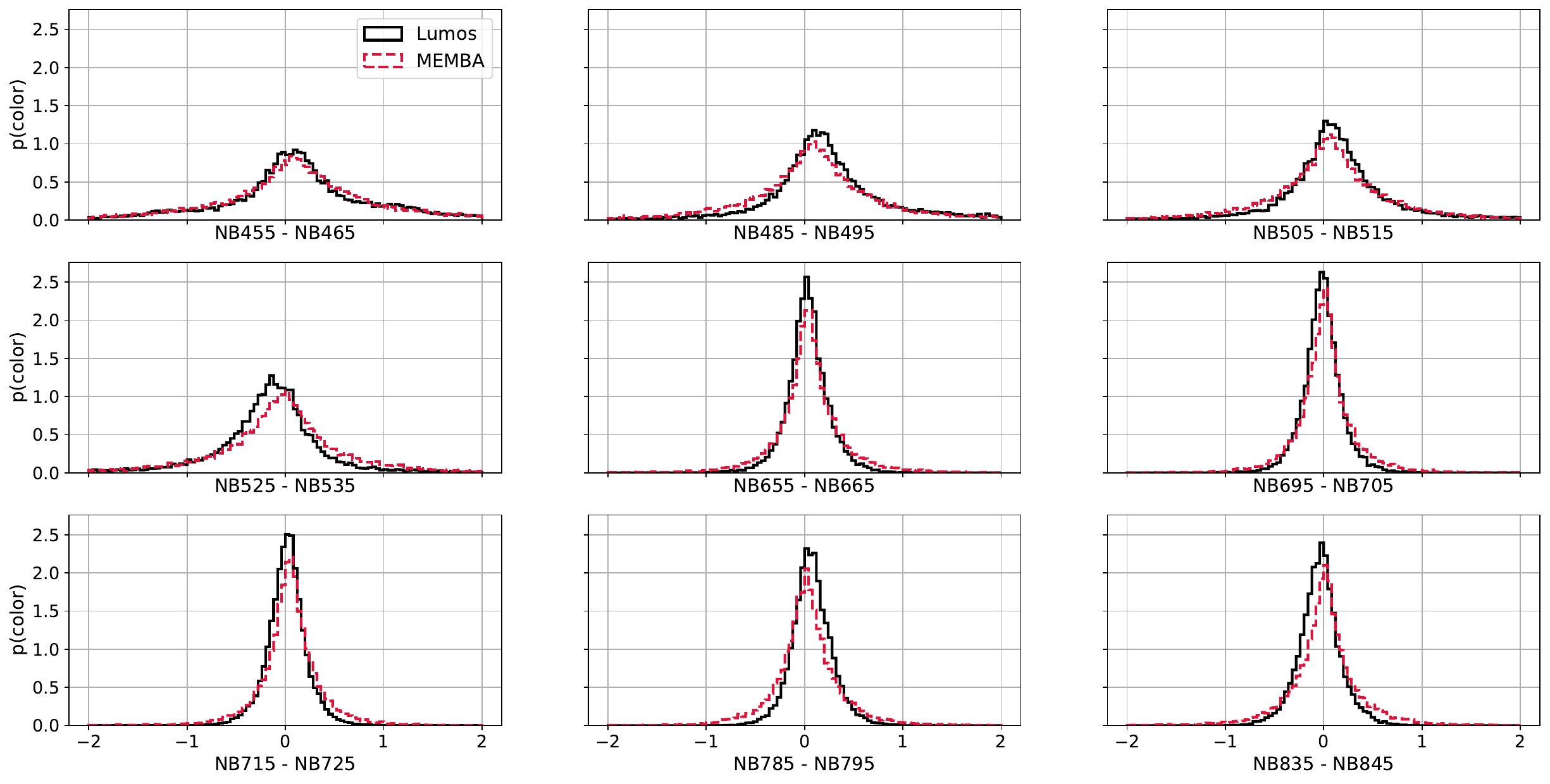}
\centering
\caption{Colour histograms for nine different narrow band filters using \textsc{Lumos} (solid black line) or \textsc{MEMBA} (dashed red line) photometries.}
\label{fig:color_hists}
\end{figure*}

Colour histograms can be used to compare different photometries of the same images. Assuming an underlying galactic colour  distribution, photometry uncertainties broaden such distribution and therefore 
the width of the colour histogram provides an idea of the photometry uncertainties. Consequently, the best photometry on a sample of galaxies is that providing narrower colour distributions. In Figure \ref{fig:colorhist} (\S \ref{sec:pausresults:singleexp}), we showed the NB785-NB795 histogram, which displayed a narrower distribution for \textsc{Lumos} than \textsc{MEMBA}. Here, we show the colour histograms results for the rest of the bands. Figure \ref{fig:color_hists} shows the colour histogram of nine different narrow bands with the \textsc{Lumos} photometry (black solid line) and the \textsc{MEMBA} photometry (red dashed line).\\

\begin{figure}
\includegraphics[width=0.47\textwidth]{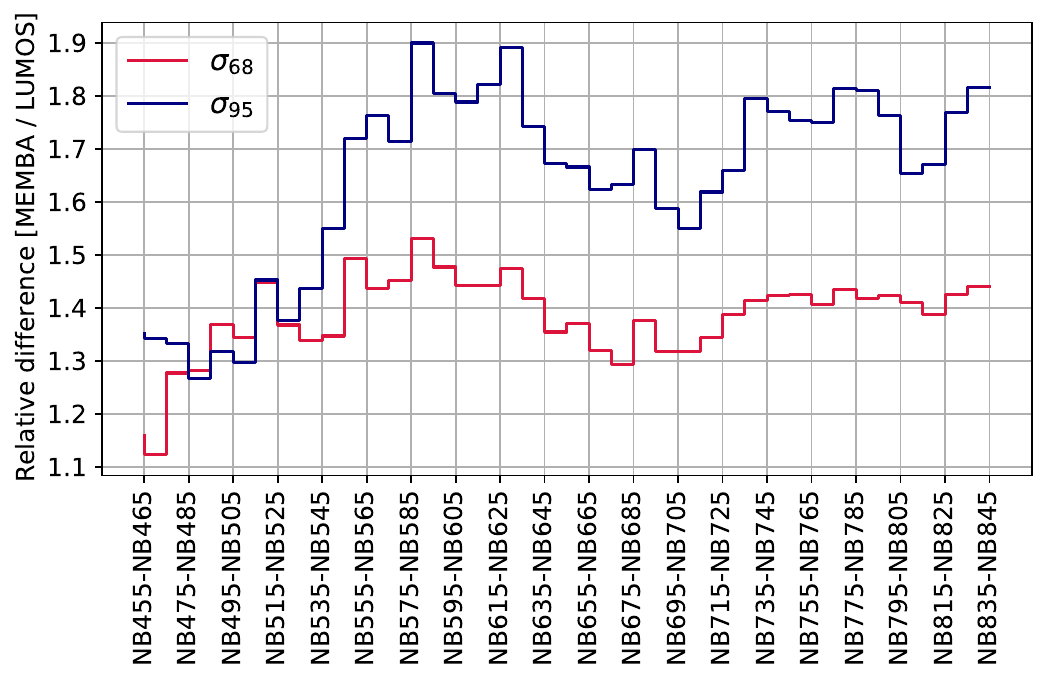}
\centering
\caption{Relative difference between the effective width of the colour histograms of the PAUS photometry with \textsc{Lumos} and \textsc{MEMBA}. The effective width has been estimated with $\sigma_{\rm 68}$ (blue line) and $\sigma_{\rm 95}$ (red line).}
\label{fig:summary_colorhists}
\end{figure}

Furthermore, Figure \ref{fig:summary_colorhists} shows the relative difference in  the effective width of the colour histograms with the photometries from \textsc{Lumos} and \textsc{MEMBA}. As in Section \ref{sec:pausresults:singleexp}, the effective widths are estimated with $\sigma_{\rm 68}$ (Eq. \ref{eq:sigma68}) and  $\sigma_{\rm 95}$, which is equivalent to  $\sigma_{\rm 68}$ but considering the 2.5 and 97.5 quantiles. \textsc{Lumos} provides narrower colour histograms in all the NB filters. The relative difference in $\sigma_{\rm 68}$ oscillates between 30\% and 40\% in all NBs but the bluest, where it is $\approx$15\%. With $\sigma_{\rm 95}$, the relative effective width is lower at the first eight bluer bands, $\approx$30\%, and increases to $\approx$70\% for the rest of the bands. The relative difference in $\sigma_{\rm 95}$ is systematically larger than  with $\sigma_{\rm 68}$, which is likely related with very noisy measurements. Photometry measurements in the tails of the colour histograms will not affect the $\sigma_{\rm 68}$ measurement, however these will be accounted in $\sigma_{\rm 95}$. Consequently, Figure \ref{fig:summary_colorhists} would be showing that the number of outlier observations is lower in \textsc{Lumos} than in \textsc{MEMBA}.

\section{Photometry and photo-z correlations with galaxy parameters}
\label{App:photometrycorrs}
\begin{figure*}
\centering
\includegraphics[width=0.49\textwidth]{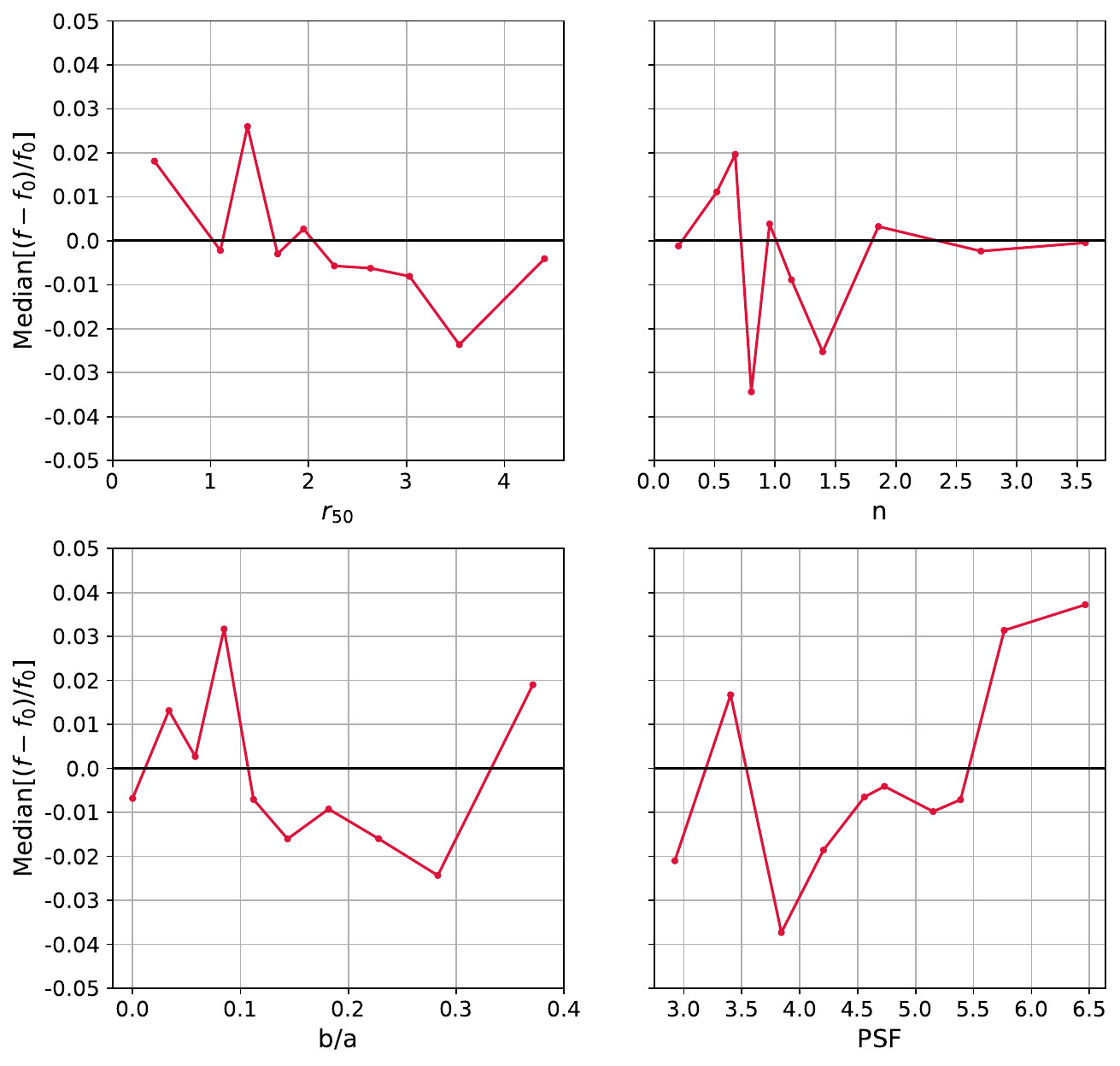}
\includegraphics[width=0.49\textwidth]{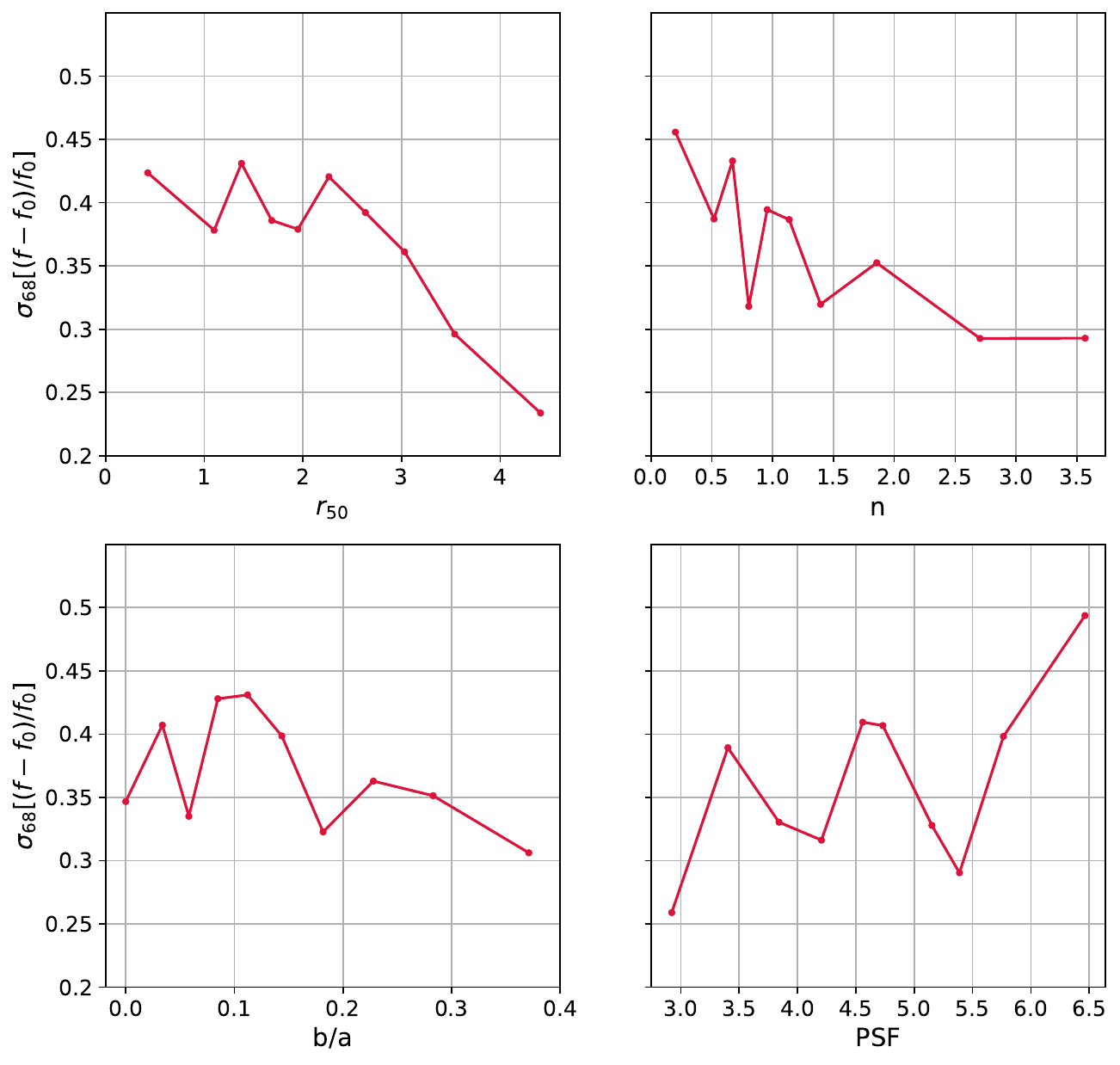}
\caption{\emph{Left: Bias in the photometry measurements as a function of the galaxy size ($r_{\rm 50}$}), galaxy shape (S\'ersic index), ellipticity (b/a) and PSF.  \emph{Right}: Precision in the photometry as a function of the same parameters.}
 \label{fig:sims_corr_params}
\end{figure*}

\begin{figure*}
\centering
\includegraphics[width=0.49\textwidth]{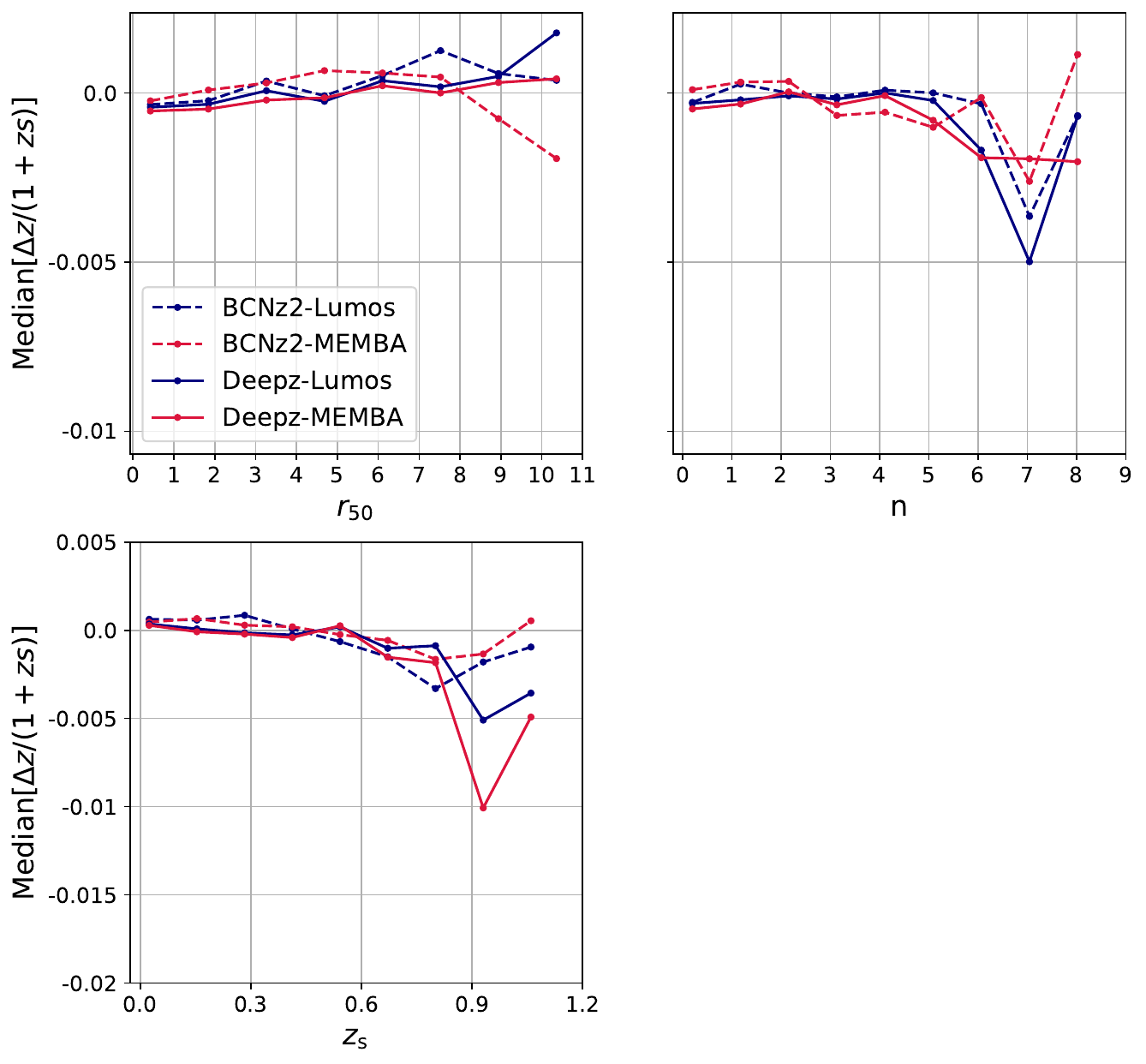}
\includegraphics[width=0.49\textwidth]{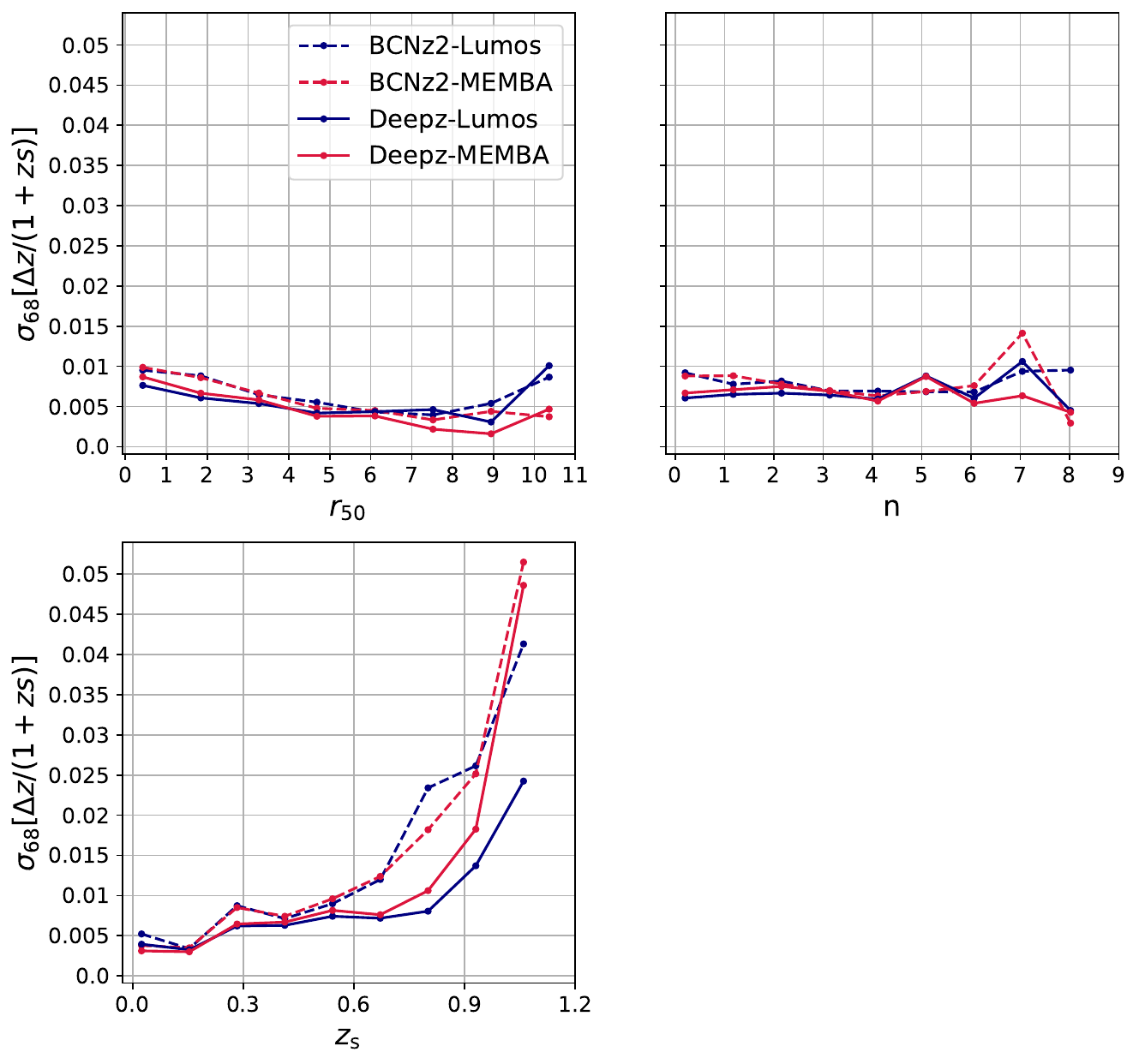}
\caption{\emph{Left}: Bias in the photo-z measurements as a function of the galaxy size ($r_{\rm 50}$}), galaxy shape (S\'ersic index) and the spectroscopic redshift ($z_{\rm z}$).  \emph{Right}: Precision in the photo-z as a function of the same parameters.
 \label{fig:zb_corr_params}
\end{figure*}

Figure \ref{fig:sims_comparison_algorithms} only showed the bias and the precision of the photometry obtained with \textsc{Lumos} (blue solid line) as a function of magnitude on simulations. In this appendix, we are extending the exploration of the photometry (\S\ref{App:photometrycorrs:photometry}) and the photo-z (\S\ref{App:photometrycorrs:photo-z}) predictions as a function of other galaxy parameters as e.g. the galaxy size or ellipticity.

\subsection{\textsc{Lumos} photometry correlation with galaxy parameters}
\label{App:photometrycorrs:photometry}
Figure \ref{fig:sims_corr_params} shows the bias and the precision of the \textsc{Lumos} photometry as a function of the galaxy size ($r_{\rm 50}$), the galaxy shape (S\'ersic index, n), the galaxy ellipticity (b/a) and the PSF, binned in ten equally populated bins. 
The photometry does not show a significant bias with any of the galaxy parameters. Note that the galaxy bias increment for higher PSF values is expected since this parameter directly correlates with the quality of the data.
The largest galaxies in the dataset tend to have slightly underestimated flux predictions ($\approx$ 1-2\%). This could be a consequence of the fixed stamp size, which could lead to a small leak of light. Nevertheless, this should have a weaker effect on \textsc{Lumos} than on other non-trainable algorithms, since the network can learn to predict reasonable fluxes when the galaxy is partially outside the stamp. \\

The photometry precision is better for larger galaxies. Furthermore, the precision is higher for larger S\'ersic indices, which is expected since larger S\'ersic indices correspond to bigger and brighter galaxies. In contrast, we do not see any correlation between the photometry precision and the galaxy ellipticity.

\subsection{Photo-z correlation with galaxy parameters}
\label{App:photometrycorrs:photo-z}
Figure \ref{fig:zb_corr_params} explores the photo-z performance with the \textsc{BCNz2} template fitting and the \textsc{Deepz} machine learning codes as a function of the galaxy size ($r_{\rm 50}$), the galaxy shape (S\'ersic index) and the spectroscopic redshift. This is presented for both the \textsc{Lumos} and the \textsc{MEMBA} photometries. In this case, these quantities are binned in 10 equally spaced bins, so that that we can explore the photo-z performance on the edges of the training set distributions.\\

Overall, the photo-z bias (first and second columns) is not affected neither by the size nor the shape of the galaxy with any of the codes or photometries. The photo-zs are also unbiased as a function of spectroscopic redshift, only presenting a $\approx$1\% bias at high redshifts with the \textsc{MEMBA} photometry and the \textsc{Deepz} code (blue solid line). Using \textsc{Deepz} on the \textsc{Lumos} photometry also presents a $\approx$0.5\% bias, while such biases disappear with the \textsc{BCNz2} algorithm on both photometries. This suggests that it might be triggered by the photo-z method.\\

The photo-z precision (third and forth rows) shows a similar trend with the galaxy size and shape using the \textsc{MEMBA} or \textsc{Lumos} photometries. Note that  \textsc{Lumos} provides better photo-z precision for small galaxies, while \textsc{MEMBA} gives better photo-zs for larger galaxies. This is potentially related to discrepancies between the training image simulations and the data (see \S\ref{sec:photoz}). Such differences  affect more large bright galaxies, as these are more resolved. A similar effect can be noted as a function of S\'ersic index.\\

The photo-z precision with spectroscopic redshift presents a similar trend for both photometries, exhibiting better photo-zs for $z_{\rm s}$>1 with both the \textsc{BCNz2} and \textsc{Deepz} codes on the \textsc{Lumos} photometry. At high redshfits, the improvement with the \textsc{Lumos} photometry and the \textsc{Deepz} code is remarkable.

\end{appendices}


\end{document}